\documentclass[%
 reprint,
 amsmath,amssymb,
 aps,
nofootinbib]{revtex4-2}

\usepackage{float}
\usepackage{amsmath,latexsym,amssymb,amsfonts}
\usepackage{subcaption}
\usepackage{dcolumn}
\usepackage{appendix}
\usepackage{mathtools}
\usepackage{graphicx}
\usepackage{relsize}
\usepackage{parallel,enumitem}
\usepackage[usenames,dvipsnames]{color}

\usepackage{dcolumn}
\usepackage{bm}
\usepackage{hyperref}
\usepackage{mathrsfs}
\hypersetup{
    colorlinks=true,
    linkcolor=blue,
    filecolor=magenta,      
    urlcolor=cyan,
    pdftitle={Overleaf Example},
    pdfpagemode=FullScreen,
    }
\usepackage{tensor}
\usepackage{hyperref} 

\begin{document}	

\title{\textbf{More Nonlinearities? \\II. A Short Guide to First- and Second-Order Electromagnetic Perturbations in the Schwarzschild background.}}

\author{Fawzi Aly${}^{1}$
}
\thanks{Corresponding author}
\email{mabbasal[AT]buffalo.edu}

\author{ Mahmoud A. Mansour${}^2$
}
\email{ mansour[AT]iis.u-tokyo.ac.jp}

\author{ Dejan Stojkovic${}^1$
}
\email{ds77[AT]buffalo.edu}
\affiliation{${}^1$HEPCOS, Physics Department, SUNY at Buffalo, Buffalo, New York, USA}
\affiliation{${}^2$Department of Physics, The University of Tokyo, Kashiwanoha, Kashiwa, Chiba 277-8574, Japan}

\begin{abstract}
We study second-order electromagnetic perturbations in the Schwarzschild background and derive the effective source terms for Regge-Wheeler equation which are quadratic in first-order gravitational and electromagnetic perturbations. In addition to the induced mixed quadratic modes, we find that linear gravitational modes are also excited, with amplitudes dependent on the electromagnetic potential. A toy model involving a Dirac delta function potential demonstrates mixing of linear gravitational and electromagnetic perturbations with frequencies \( \omega^{(1)}  \) and \( \Omega^{(1)} \), resulting in the second-order QNM mixing in the electromagnetic field at \( \Omega^{(2)} =\Omega^{(1)} + \omega^{(1)}  \). This complements prior work in \cite{aly2024nonlinearities} on the second-order gravitational perturbation mixing and highlights potential applications in multi-messenger astrophysics for systems observed by LIGO-Virgo-KAGRA (LVK) and upcoming LISA. We also study first-order perturbations due to a point charge and show it could be reduced to a one-dimensional path integral. Within the toy model, we investigate the first-order electromagnetic perturbation due to a radially free-falling single charge \( q \) and radial dipole moment \( p = q \eta \), employing semi-analytical and numerical methods. For the dipole case, we show that the QNM perturbation is excited with a nearly constant amplitude. 
Future work will focus on incorporating mixing in more realistic potentials and exploring numerical approach in the context of rotating spacetimes.
\end{abstract}

\maketitle

\section*{Introduction}
The study of quasinormal modes (QNMs) is fundamental to understanding how black holes (BHs) respond to perturbations. In general relativity (GR), QNMs play an even more critical role, as they are only parametrized by BH macroscopic properties: mass \( M \), angular momentum \( J \), and charge \( Q \)—making them characteristic signatures of these enigmatic objects \cite{ReviewArticle_QNM_Berti_Cardoso,Kokkotas_1999,konoplya2011quasinormal,ferrari2008quasinormal}.  In this context, the black hole spectroscopy program was established. While observing a single QNM provides a direct measurement of these parameters, detecting two or more QNMs offers a valuable consistency check. This enables testing the validity limits of black hole perturbation theory (BHPT) in modeling the late-stage ringdown phase \cite{ReviewArticle_QNM_Berti_Cardoso,2Modes_As_test_Per_theory_1,2Modes_As_test_Per_theory_2_Agnosticspectroscopy}.
\vspace{1mm}

\noindent While gravitational QNMs have garnered significant attention—particularly following the groundbreaking gravitational wave (GW) detections by (LIGO-Virgo-KAGRA) LVK \cite{GW150914,GW200105_GW200115}—electromagnetic QNMs remain a relatively under-explored area with intriguing theoretical implications, and potential astrophysical relevance. This is particularly true in studying neutron stars (NSs), and charged BH properties. Electromagnetic QNMs encode some information about how electromagnetic fields evolve and decay around vacuum BH spacetimes, for example like those considered as perturbations of   Kerr type remnants after magnetized mergers \cite{Cardoso_EM_Formulation_Anti_de_Sitter}. Understanding these modes is essential for better interpretation of electromagnetic radiation spectrum emitted during astrophysical events, including black hole-neutron star (NSBH) and binary neutron star (BNS) mergers, gravitational collapse of unstable magnetars, magnetar flares, and accretion processes near BHs.
\vspace{1mm}

\noindent In the context of multi-messenger astronomy \cite{multi_messenger,Abbott_2017__,Margalit_2017,Branchesi_2016}, the simultaneous detection of gravitational and electromagnetic signals offers a unique opportunity to probe the strong-field regime and more thoroughly examine astrophysical events where electromagnetic effects are switched on. The gravitational and electromagnetic signals are expected to provide both complementary and overlapping information about the same astrophysical event, allowing for cross-checks. Although, in principle, understanding electromagnetic QNM spectra might lead to more accurate extraction of information in the aforementioned systems \cite{andersson1996quasinormal, ferrari2008quasinormal} and enable further testing of GR and its coupling with electromagnetism \cite{aly2024nonlinearities}, the state of the art in any realistic signal processing for BNS is to ignore the QNM-component. This is partially due to the complexity of magnetized mergers and lack of full description of how electromagnetic field behaves in the vicinity of the ejecta and in plasma environment.
\vspace{1mm}

\vspace{1mm}


\noindent \noindent In the context of numerical simulations, scenarios involving mergers of charged binary BHs or BNS, magnetar collapse, and NSBH systems without significant disruption offer a higher probability of resolving electromagnetic QNM behavior. Current and future simulations of charged or magnetized systems could shed light on the remnant's ringdown for these scenarios, especially following recent attempts to identify electromagnetic QNMs, (i.e. in  \cite{Most_2018,Most_2024}). Given that magnetic fields in astrophysical BNS mergers can reach up to \( B \sim 10^{12} \, \text{T} \) \cite{abbott2017gw170817}, it is intriguing to explore the role of electromagnetic fields in these extreme conditions.
\vspace{1mm}

\noindent In \cite{aly2024nonlinearities}, we explored whether electromagnetic fields can leave imprints on gravitational radiation, focusing more on extreme mass ratio inspiral (EMRI) scenarios \cite{Aly_2023,TestingGravityWithEMRIs,Amaro_Seoane_2023,Berti_2006}. We reported that it was plausible for electromagnetic effects to compete with the second-order gravitational perturbations in a NSBH merger. With a lower limit estimate for astrophysical NS of mass \( 2 M_\odot \), and with a magnetic field of order \( \sim 10^{7} \, \text{T} \), in an inspiral around a supermassive black hole of mass \( 10^6 M_\odot \). However, in the context of stellar-mass mergers, the situation is more challenging in the absence of a proper way to define perturbation parameters, so perhaps further analytical and numerical investigations are needed.

\vspace{1mm}

\noindent However, since gravitational effects are expected to dominate in astrophysics, this complementary work aims to generically investigate the reverse question: \textit{Can we detect gravitational signatures in the electromagnetic radiation?} Such phenomena should be important in analyzing and modeling electromagnetic counterparts of magnetized collapse or mergers in future.
\vspace{1mm}


\noindent The work in this paper addresses several questions. First, we provide a brief review of the Regge-Wheeler formalism for electromagnetic perturbations. We then study second-order electromagnetic perturbations, where the electromagnetic-gravitational interaction, with its first-order modes, acts as an effective source term. We solve this problem for a sample source term for the purpose of illustration. Following this, we go back to first order perturbation, and solve for the electromagnetic perturbations induced by a point charge and also an ideal dipole, after approximating the Regge-Wheeler potential with a Dirac delta function. The choice of an ideal dipole serves as an analogy for astrophysical systems with relatively small macroscopic charges. Finally, we report our numerical solutions for the perturbation due to the dipole.
\vspace{1mm}

\noindent The structure of this paper is as follows. In Section \ref{Preliminary}, we  outline the electromagnetic perturbation in Schwarzschild coordinates. Next, in Section \ref{Second order Electromagnetic Perturbation}, we formulate the second-order perturbation and derive the source terms for the second-order master Regge-Wheeler formalism for spin \( s = 1 \). By the end of this section, we solve for a sample source using the Dirac delta function approximation to illustrate electromagnetic-gravitational mixing, similar to \cite{aly2024nonlinearities}. Then, we return to first-order perturbations in Section \ref{Perturbation Due to Point Charges} and show that, in the time domain, the inhomogeneous solution due to a point charge can be reduced to a one-dimensional path integral, and we discuss several examples. In Sections \ref{Dirac Delta potaioal} and \ref{First-order Electromagnetic Perturbation Due to Point Charges}, we investigate the first-order electromagnetic perturbation due to a radially free-falling point charge and then a radial dipole in the Dirac delta approximation using a semi-analytical approach. Next, we also solve for the perturbation due to the dipole numerically and comment on the solution in Section \ref{Numerical Solution}. Finally, we discuss the results of our analysis and potential directions for future work in Section \ref{Conclusion}. Throughout this work, we use natural units where \( G = c = 1 \).

\section{Preliminary}\label{Preliminary}
In this section, we briefly review the spherical decomposition of Maxwell's equations on a spherically symmetric spacetime, such as Schwarzschild. We discuss how the Faraday tensor $F$ can be expressed in terms of perturbation quantities $\Phi_{lm}$, as well as the current multipoles $j_{lm}$, and then obtain the corresponding electromagnetic stress-energy tensor $T$. These few steps will allow us to solve the toy model introduced in \ref{Dirac Delta potaioal}.
\subsection{Maxwell equations}\label{Maxwell Equations on Schwarzschild Background}
To derive the Regge-Wheeler equations for electromagnetic perturbations in a Schwarzschild spacetime \cite{Cardoso_EM_Formulation_Anti_de_Sitter,KarlMartel_2005_RW,mARTELphdthesis,Gerlach_Ulrich}, the vector potential \(A_{\mu}\) and the current vector \(J_{\mu}\) are decomposed into even and odd components, each of which is expanded in terms of 4-vector spherical harmonics, as outlined in Appendix \ref{Vector Harmonics}. Consequently, the Faraday tensor \(F_{\mu\nu}\), and therefore Maxwell's equations, are similarly divided into sectors of even and odd parity. The first order even and odd electromagnetic perturbations 
\begin{equation}
\begin{aligned}
&{}_{e}\Phi^{(1)}_{lm}(t, x)=\frac{r^2}{\lambda}\left(\frac{\partial h_{lm}}{\partial t}-\frac{\partial g_{lm}}{\partial r} \right)\\
&{}_{o}\Phi^{(1)}_{lm}(t, x)= a_{lm}
\end{aligned}
\end{equation}
are governed by the same potential \( {}_{s=1}V_{RW}=\frac{\lambda f(r)}{r^2} \), where \(\lambda = l(l+1)\) and \(l\) represents the angular number, but differ in how the source term \( {}_{J} \mathcal{S}^{(1)}_{o,e \, lm}(t, x) \) is derived from $J_{\mu}$. The governing equation for these perturbations is expressed as:
\begin{equation}\label{EM RW equation}
\begin{aligned}
    &\mathcal{L}_{EM} := \partial_{xx} - \partial_{tt} - \, {}_{s=1}V_{RW}(x), \\
    &\mathcal{L}_{EM} \, {}_{o,e}\Phi^{(1)}_{lm}(t, x) = {}_{J} \mathcal{S}^{(1)}_{o,e \, lm}(t, x),
\end{aligned}
\end{equation}
where \(x\) is the tortoise radial coordinate, related to the areal radial coordinate through \( dr = f(r) dx \), while \(t\) is the time measured by a clock asymptotically far from the Schwarzschild black hole. In addition, \( {}_{J} \mathcal{S}^{(1)}_{o,e \, lm}(t, x) \) is formulated in terms of the current vector multipoles \( j_{(i)\, lm} \),where $i=1,2,3,4$, with an example provided in section \ref{Perturbation Due to Point Charges}. 
\vspace{1mm}

\noindent This ODE governing the perturbations has two regular singularities (rank = 1) at \(r = 0\) and \(r = 2M\), along with an irregular singularity (rank = 2) as \(r \rightarrow \infty\). This structure classifies it as a confluent Heun ODE \cite{ronveaux_2007,Aly_2023}, which can be solved using confluent Heun functions $H_C$ or, in some cases, a narrower subset of confluent Heun polynomials \cite{borissov2010exact}. The QNMs boundary conditions are imposed on the solution, and together with two initial conditions, the full solution can be specified, the reader can find all the details in  Section II.A of \cite{aly2024nonlinearities}.

\subsection{Faraday tensor}\label{Faraday Tensor on Schwarzschild Background}
After successfully separating and decoupling the system into its two degrees of freedom ${}_{o,e}\Phi^{(1)}_{lm}$, the Faraday tensor \(F_{\mu \nu}\) can now be expressed in terms  ${}_{o,e}\Phi^{(1)}_{lm}$ and the current multipoles \( j_{(i) \, lm}(t, x) \) for $l>0$. However, before proceeding, let's first address the monopole term $l=0$ of \(F_{\mu\nu}\).

\subsubsection{Monopole Case}\label{Monopole Case}
Given the multipole decomposition, the monopole case can be solved separately, as the potential vanishes. The coupled equations for the scalar field $\Phi_{00}$ are:
\begin{equation}\label{monopole system of eqs}
\begin{aligned}
\frac{\partial \Phi_{00}}{\partial r} &= \frac{r^2 \, j_{(1)\, 00}}{f(r)} \\
\frac{\partial \Phi_{00}}{\partial t} &= r^2 f(r) \, j_{(2)\, 00} .
\end{aligned}
\end{equation}

When the perturbation is due to a collection of discrete charges, the solution reduces to a summation over the individual charges \( q_i \), each multiplied by a step function dependent on their \( r - r_{p_i}(t) \) argument
\begin{equation}
\Phi_{00} = \sum_{i} q_i \, \Theta[r - r_{p_i}(t)],
\end{equation}
where \( \Theta \) denotes the Heaviside step function and \( r_{p_i}(t) \) is the radial coordinate of the \( i \)-th charge parameterized by \( t \). This solution should resemble a static perturbation in the background spacetime at radii greater than the position of the charge distribution. Such a perturbation effectively charges the spacetime, transforming the region outside the charge distribution into a member of the Reissner–Nordström family of spacetimes\cite{chandrasekhar_1983,griffiths_podolský_2012,Aly_2024}:
\begin{equation}
F_{\mu\nu00}=
\begin{pmatrix}
0 & \frac{Q_{\text{tot}}(t,r)}{r^2} & 0 & 0 \\
** & 0 & 0 & 0 \\
0 & 0 & 0 & 0 \\
0 & 0 & 0 & 0
\end{pmatrix}
\end{equation}
where \( Q_{\text{tot}} = \sum_{i} q_i \, \Theta[r - r_{p_i}(t)] \), somewhat analogous to Gauss's law in electrostatics in the exterior region. It is important to note that if the source includes no monopole contribution, this perturbation vanishes, \textit{as in the case of dipole moments} when modeling astrophysical systems with relatively small macroscopic charge relative to the higher moments. Subsequently, any electromagnetic moments would be radiated away on a time scale comparable to that of the merger and post-merger phases \footnote{We are assuming that the remnant spacetime exterior is vacuum. However, if there were accretion disks or plasma present, the situation might differ, and the same conclusions may not hold.}. During the process when the BHPT is expected to hold, the ${}_{o,e}\Phi^{(1)}_{lm}$ should describe this radiation. Furthermore, if the electromagnetic moments are sufficiently large, this radiation could be substantial, albeit subdominant to gravitational radiation. In such a case, coupling effects between the two radiations must be considered, which is the purpose of this work and its companion \cite{aly2024nonlinearities}.

\subsubsection{Multipoles $l>0$}\label{Multipoles}
 For higher multipoles \( l > 0\), the odd components of the Maxwell tensor ${}_{o} F_{\mu\nu lm}$ in the multipole expansion take the following forms for a generic source term:
\begin{equation}
{}_{o} F_{lm}=
\begin{pmatrix}
0 & 0 & -\csc\theta Y_{lm}' \frac{\partial {}_o \Phi_{lm}}{\partial t} & -\csc\theta Y_{lm}' \frac{\partial {}_o \Phi_{lm}}{\partial r} \\
0 & 0 & \sin\theta \dot{Y}_{lm} \frac{\partial {}_o \Phi_{lm}}{\partial t} & \sin\theta \dot{Y}_{lm} \frac{\partial {}_o \Phi_{lm}}{\partial r} \\
** & ** & 0 & -\lambda \, \sin\theta Y_{lm} \, {}_o \Phi_{lm} \\
** & ** & ** & 0
\end{pmatrix}.
\end{equation}

The even part of the Maxwell tensor could be divided into two contributions, which can be expressed as:
\[
{}_{e} F_{\mu\nu lm} = {}_{e1} F_{\mu\nu lm} + {}_{e2} F_{\mu\nu lm},
\]
where the first contribution, \( {}_{e1} F_{\mu\nu lm} \), is written in terms of the even scalar perturbation \({}_{e} \Phi \), while the second contribution, \( {}_{e2} F_{\mu\nu lm} \), is expressed in terms of the $j_{(1)\, lm}$ and $j_{(2)\, lm}$ multipole components of the even current vector. 

\begin{equation}
{}_{e1} F_{lm}=
\begin{pmatrix}
0 &  \frac{\lambda}{r^2} Y_{lm} {}_{e} \Phi_{lm} & \dot{Y}_{lm} f \frac{\partial {}_{e} \Phi_{lm}}{\partial r} & Y'_{lm} f \frac{\partial {}_{e} \Phi_{lm}}{\partial r} \\
** & 0 & \dot{Y}_{lm} \frac{\partial {}_{e} \Phi_{lm}}{\partial t} \frac{1}{f} & Y'_{lm} \frac{\partial {}_{e} \Phi_{lm}}{\partial t} \frac{1}{f} \\
** & ** & 0 & 0 \\
** & ** & ** & 0
\end{pmatrix}
,
\end{equation}
and,
\begin{equation}
{}_{e2} F_{lm}=
\begin{pmatrix}
0 & 0 & -\dot{Y}_{lm} \frac{r^2 j_{(1)lm}}{\lambda} & -Y'_{lm} \frac{r^2 j_{(1)lm}}{\lambda} \\
** & 0 & -\dot{Y}_{lm} \frac{r^2 j_{(2)lm}}{\lambda} & -Y'_{lm} \frac{r^2 j_{(2)lm}}{\lambda} \\
** & ** & 0 & 0 \\
** & ** & ** & 0
\end{pmatrix}.
\end{equation}
Finally we can express the Maxwell tensor $F_{\mu\nu}$ as 
\begin{equation}\label{EM energy tesnor multiples}
\begin{aligned}
F_{\mu\nu} &= F_{\mu\nu\, 00} + \sum_{l=1}^{\infty} \sum_{m=-l}^{l} {}_{e2} F_{\mu\nu\, lm}\\ 
&+\sum_{l=1}^{\infty} \sum_{m=-l}^{l} \left({}_{e1} F_{\mu\nu\, lm} + {}_{o} F_{\mu\nu\, lm}\right)
\end{aligned}
\end{equation}
So far, we haven't made any specific assumptions about the perturbation source, yet we can make some general statements about the behavior of the Maxwell tensor in its current form.
\begin{itemize}
    
    \item While solving the system (\ref{monopole system of eqs}), it becomes clear that the monopole \(\Phi_{00}\) neither exhibits radiative behavior nor satisfies the typical boundary conditions imposed on QNMs. As \( t \rightarrow \infty \) on the clock of an asymptotic observer, the charge will freeze near the horizon, all radiative components (QNMs,tails,..etc) will dissipate. Ultimately, the system will asymptotically approach the Reissner-Nordström solution.

    \item The term \({}_{e2} F_{\mu\nu\, lm}\) contains more information about the behavior of the electromagnetic field in the neighborhood of the charge, typically localized if the source $J_{\mu}$ behaves as expected in astrophysical cases. Consequently, when calculating the stress-energy tensor \(T_{\mu \nu}\), this component needs to be regularized for discrete charge distributions to avoid infinite energy density. We shall refer to \({}_{e2} F_{\mu\nu\, lm}\) as ``internal" field from now on.
    
    \item The components \({}_{e1} F_{\mu\nu\, lm}\) and \({}_{o} F_{\mu\nu\, lm}\) are radiative in nature and represent the QNM behavior of the electromagnetic field. These components are obtained by solving the inhomogeneous Regge-Wheeler equations (\ref{EM RW equation}), where \(j_{(1)\, lm}\) and \(j_{(2)\, lm}\) source the even component, and \(j_{(4)\, lm}\) sources the odd component. The initial conditions on the Electromagnetic field will add non-trivial constraints on ${}_{o,e}\Phi^{(1)}_{lm}$, this will be highly relevant if we are focusing more on stellar-mass mergers (i.e. NSBH or BNS mergers) ringdown phase.
\end{itemize}

\subsection{Stress-energy tensor}
Given the Faraday tensor \( F_{\mu\nu} \), we can proceed to calculate the electromagnetic stress-energy tensor \( T_{EM}^{\mu\nu} \), which describes the distribution and flow of electromagnetic energy and momentum in spacetime. In spherical coordinates, \( T_{EM}^{\mu\nu} \) takes the following standard form:

\begin{equation}\label{generalcasting EM energy tensor}
T_{EM}^{\mu\nu} =
\begin{pmatrix}
\epsilon & S_r & S_\theta & S_\phi \\
S_r & P_r & \sigma_{r\theta} & \sigma_{r\phi} \\
S_\theta & \sigma_{r\theta} & P_\theta & \sigma_{\theta\phi} \\
S_\phi & \sigma_{r\phi} & \sigma_{\theta\phi} & P_\phi, \\
\end{pmatrix}
\end{equation}

\noindent Assuming that the components of the stress-energy tensor retain their familiar physical interpretations as they do in Minkowski spacetime, \( \epsilon \) represents the energy density or electromagnetic energy per unit volume; \( S_r \), \( S_\theta \), and \( S_\phi \) are the components of the Poynting vector, indicating the flow of electromagnetic energy in the radial, polar, and azimuthal directions, respectively; \( P_r \), \( P_\theta \), and \( P_\phi \) correspond to the pressure in these directions; and the off-diagonal elements \( \sigma_{r\theta} \), \( \sigma_{r\phi} \), and \( \sigma_{\theta\phi} \) describe the shear stresses in the electromagnetic field.
\vspace{1mm}

\noindent As discussed earlier, we will exclude the contribution from \({}_{e1} F_{\mu\nu\, lm}\) when calculating \( T_{EM}^{\mu\nu} \). Since the full expressions are lengthy, for completeness and illustration purposes, we will just write down the expressions for \( \epsilon \) and \( S_r \) while rest will be provided in a companion Mathematica Notebook:

\begin{equation}
\begin{aligned}
\epsilon &= \sum_{l,m}\sum_{l',m'}
\frac{A_{lm l'm'}}{2 r^2 f(r)} \bigg[
{}_{e}\Phi_{lm} \frac{\partial {}_{o}\Phi_{l'm'}}{\partial r}
- \frac{\partial {}_{e}\Phi_{lm}}{\partial r} {}_{o}\Phi_{l'm'} \\
&\quad + {}_{o}\Phi_{lm} \frac{\partial {}_{e}\Phi_{l'm'}}{\partial r}
+ \frac{\partial {}_{o}\Phi_{lm}}{\partial r} {}_{e}\Phi_{l'm'}
\bigg] \\
&+ \frac{B_{lm l'm'}}{2 r^2} \bigg[- \frac{\partial {}_{o}\Phi_{lm}}{\partial r} \frac{\partial {}_{o}\Phi_{l'm'}}{\partial r}
- \frac{\partial {}_{e}\Phi_{lm}}{\partial r} \frac{\partial {}_{e}\Phi_{l'm'}}{\partial r} \\
&\quad - \frac{1}{f(r)^2} \frac{\partial {}_{o}\Phi_{lm}}{\partial t} \frac{\partial {}_{o}\Phi_{l'm'}}{\partial t}
-\frac{1}{f(r)^2} \frac{\partial {}_{e}\Phi_{lm}}{\partial t} \frac{\partial {}_{e}\Phi_{l'm'}}{\partial t}
\bigg] \\
&- \frac{\lambda^2 Y_{lm} Y_{l'm'} \left({}_{e}\Phi_{lm} {}_{e}\Phi_{l'm'} + {}_{o}\Phi_{lm} {}_{o}\Phi_{l'm'}\right)}{2 r^4 f(r)} \\
&-\frac{Q_{tot}}{{r^4 f(r)}}\sum_{l,m} \lambda Y_{lm} {}_{e}\Phi_{lm}
- \frac{Q_{tot}^2}{2 r^4 f(r)}
\end{aligned}
\end{equation}
and 
\begin{equation}
\begin{aligned}
S_r &= \frac{1}{r^2} \sum_{l,m} \sum_{l',m'} \bigg\{
A_{lm l'm'} \left[ f(r) \frac{\partial {}_{e}\Phi_{lm}}{\partial r} 
\frac{\partial {}_{o}\Phi_{l'm'}}{\partial r} \right. \\
&\left. -\frac{1}{f(r)} \frac{\partial {}_{o}\Phi_{lm}}{\partial t} 
\frac{\partial {}_{e}\Phi_{l'm'}}{\partial t} \right]
+ B_{lm l'm'} \left[ \frac{\partial {}_{o}\Phi_{l'm'}}{\partial r} 
\frac{\partial {}_{o}\Phi_{lm}}{\partial t} \right. \\
&\left. + \frac{\partial {}_{e}\Phi_{l'm'}}{\partial t} 
\frac{\partial {}_{e}\Phi_{lm}}{\partial r} \right] 
\bigg\}
\end{aligned}
\end{equation}
where
\begin{equation}
\begin{gathered} 
A_{lm l'm'} = \csc\theta 
\left[
\dot{Y}_{lm} Y'_{l'm'} 
- Y'_{lm} \dot{Y}_{l'm'}
\right],\\
B_{lm l'm'} = \csc^2\theta 
\left[
Y'_{lm} Y'_{l'm'} 
+ \dot{Y}_{lm} \dot{Y}_{l'm'}
\right]
\end{gathered}
\end{equation}
 The main takeaways from this brief summary so far are as follows:
\begin{itemize}
    \item The stress-energy tensor \( T_{EM}^{\mu\nu} \) involves both linear terms in \({}_{e}\Phi_{lm}\) and \({}_{o}\Phi_{lm}\), as well as quadratic terms involving products of the form \({}_{e}\Phi_{lm} \times {}_{e}\Phi_{l'm'}\), \({}_{o}\Phi_{lm} \times {}_{o}\Phi_{l'm'}\), and \({}_{e}\Phi_{lm} \times {}_{o}\Phi_{l'm'}\) and their derivatives, summed over the indices \(l\), \(l'\), \(m\), and \(m'\). As a result, we can anticipate that the frequencies associated with the mixing modes will be either linear or quadratic as argued for in \cite{aly2024nonlinearities}.
    
    \item We can decompose equation (\ref{generalcasting EM energy tensor}) using, for example, the 4D tensor harmonics defined as in \cite{aly2024nonlinearities}. In the Laplace spectral space (or less formally in the Fourier space), this will allow us to study the spectral decomposition of the energy flux, shear, and density as functions of \(r\) for each multipole \(lm\) and frequency $\Omega_n$ similar to \cite{Tiomno_1972dq,Ruffini_1972uh}. Furthermore, we can examine how they couple to other radiative multipoles \(l'm'\) of \(\Phi_{l'm'}\).
    
    \item The spherical decomposition of \( T_{EM}^{\mu\nu} \) in expression (\ref{EM energy tesnor multiples}) is not fully compatible with gravitational perturbations. To resolve this, we propose projecting it onto the tensorial spherical harmonics provided in Appendix A of \cite{aly2024nonlinearities} or following ref \cite{KarlMartel_2005_RW} as  demonstrated in Appendix C,D,E, and F in \cite{aly2024nonlinearities}, ensuring compatibility with the Regge-Wheeler gravitational analysis\cite{aly2024nonlinearities}.
    
    \item As \(t \rightarrow \infty\), both \(\Phi\) and its derivatives will vanish, and the total charge \(Q_{\text{tot}}\) will appear frozen near the horizon from the perspective of an asymptotic observer. Consequently, only the monopole term, i.e., the Reissner–Nordström charge, will be measurable at ``late times". However, as different radiative modes (i.e. QNMs and tails) have different decaying  rates, that suggest the we should be careful what time-scale we are interested in studying.
\end{itemize}

\begin{figure}[H]
\includegraphics[width=8.5cm]{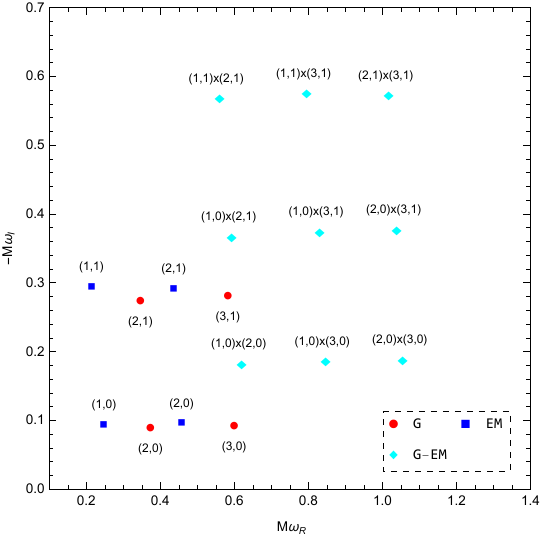}
\captionsetup{format=plain,justification=justified}
\caption{ Comparison of gravitational and electromagnetic QNMs in Schwarzschild spacetime. Gravitational frequencies \( \omega^{(1)}_{ln} \) for \( l = 2, 3 \) and \( n = 0, 1 \) are compared with electromagnetic frequencies \( \Omega^{(1)}_{ln} \) for \( l = 1, 2 \) and \( n = 0, 1 \). For the quadratic QNMs, the real part of the frequencies is given by the sum 
        \(
        \Omega^{(2)}_{(i \times j)R} = \Omega^{(1)}_{(i)R} \pm \omega^{(1)}_{(j)R}      \)  of the linear modes, while the imaginary part, representing the reciprocal of the decay time, is given by  \(  \Omega^{(2)}_{(i \times j)I} = \Omega^{(1)}_{(i)I} + \omega^{(1)}_{(j)I}  \) as a sum of the corresponding linear terms. The blue squares and red dots represent the electromagnetic and gravitational QNMs, respectively, while cyan rhombuses indicate the quadratic QNMs.
     }
\label{fig:myPlot}
\end{figure} 
\section{Second order electromagnetic perturbation}\label{Second order Electromagnetic Perturbation}

In \cite{aly2024nonlinearities}, the coupling between electromagnetic and gravitational modes was investigated perturbatively in the framework of minimally coupling Einstein-Maxwell system. It was reported that, in the simplest case of mixing between the two fields, when the electromagnetic field sources gravitational perturbations, it excites gravitational modes that are linear and quadratic in the electromagnetic QNMs. Also amplitude of the linear gravitational QNMs get correction from electromagnetic field. As we go higher in mixing, more coupling between the modes is expected to take place, such that modes quadratic in both fields arise.
\vspace{1mm}

\noindent On the other hand, at first and second order in $\alpha$, the Maxwell equation for the perturbed electromagnetic field is expressed as:
\begin{equation}
\begin{gathered}
\alpha \nabla_\mu^{(0)} F^{(1)\mu \nu}=\alpha J^{(1)\nu}\\
\alpha^2 \nabla_\mu^{(0)} F^{(2)\mu \nu}=\alpha^2 J_{eff}^{(2) \, \nu} ,
\end{gathered}
\end{equation}
where $F^{(1,2)}_{\mu\nu}$ and $J^{(1,2)\nu}$ represent the first and second -order perturbation of the electromagnetic field and second-order current source term respectively. The covariant derivative $\nabla_\mu^{(0)}$ is taken with respect to the unperturbed Schwarzschild metric $g_{\mu\nu}^{(0)}$, hence it's straightforward to reduce the second order field perturbation to their corresponding Regge-Wheeler second odd \({}_{o}\Phi^{(2)}_{lm}\) and even \({}_{e}\Phi^{(2)}_{lm}\) perturbation:
\[
 \mathcal{L}_{EM} \, {}_{o,e}\Phi^{(2)}_{lm}(t, x) = {}_{\text{eff}} \mathcal{S}^{(2)}_{o,e \, lm}(t, x) .
\]

At the second order, there is additional effective nonlinear contribution which  $J_{\text{eff}}^{(2)\nu}$ represents coupling between the first-order electromagnetic field $F^{(1)}_{\mu\nu}$ and the first-order metric perturbation $g^{(1)}_{\mu\nu}$. The effective second-order current, $J_{\text{eff}}^{(2)\nu}$, is given by
\begin{equation}
\begin{aligned}
J_{\text{eff}}^{(2)\nu} &= -\frac{1}{2} \left( g^{(0)\gamma \mu} F^{(1)\alpha\nu} + g^{(0)\nu \gamma} F^{(1)\mu \alpha} \right) \\
&\times \left( \nabla_\mu^{(0)} g_{\alpha \gamma}^{(1)} + \nabla_\alpha^{(0)} g_{\mu \gamma}^{(1)} - \nabla_\gamma^{(0)} g_{\alpha \mu}^{(1)} \right),\\
&=\frac{1}{2} F^{(1)\nu \alpha}  \nabla_\alpha^{(0)} g_{\mu}^{\mu\,(1)}
\end{aligned}
\end{equation}
It is straightforward to find the corresponding Regge-Wheeler sources corresponding to $J_{\text{eff}}^{(2)\nu}$, through using vector spherical harmonics in Appendix \ref{Vector Harmonics}. We report the final result for even and odd sources respectively here
\begin{widetext}
\begin{equation}
\begin{gathered}
{}_{\text{eff}} S_{e \, lm} = \frac{1}{2\lambda} \left( \frac{\partial^2}{\partial t^2}-\frac{\partial^2}{\partial x^2}\right)  \tilde{j}_{lm} \\
\tilde{j}_{lm} = \left[ -2 \left(  \frac{\partial}{\partial x} + \bar{A} \right) \psi_{e \, l' m'} \right]
\left[ Q_{\text{tot}} \, \delta_{l l'} \delta_{m m'} + (-1)^m \, \lambda'' \, \Phi_{e \, l'' m''} \, B_{l, -m; l', m', l'', m''} \right], \\
\bar{A} = \frac{2}{r( \lambda - 2) + 6M} \left[ \left( \frac{1}{2} \lambda - 1 \right) \frac{1}{2} \lambda + \frac{3M}{r} \left( \frac{1}{2} \lambda -f \right) \right].
\end{gathered}
\end{equation}

\begin{equation}
\begin{gathered}
{}_{\text{eff}} S_{o \, lm} = (-1)^m \frac{f}{2} \Bigg\{ - C_{l, -m; l', m', l'', m''} \frac{\lambda'' \Phi_{o \, l'' m''}}{r^2} \left[ -2 \left(  \frac{\partial}{\partial x} + \bar{A} \right) \psi_{e \, l' m'} \right],\\ + g^{aa} \frac{\partial}{\partial x_a} \left[ -2 \left(  \frac{\partial}{\partial x} + \bar{A} \right) \psi_{e \, l' m'} \right] \Bigg[ C_{l, -m; l'', m'', l', m'} \frac{\partial \Phi_{o \, l'' m''}}{\partial x_a}
+ D_{l, -m; l'', m'', l', m'}, 
\left( \epsilon_{a}^{b} \frac{\partial \Phi_{e \, l'' m''}}{\partial x_b} + \frac{r^2 j_{(a) \, l'' m''}}{\lambda''} \right) \Bigg] \Bigg\}\\
\epsilon_{a}^{b} = \begin{pmatrix} 0 & f \\ \frac{1}{f} & 0 \end{pmatrix}.
\end{gathered}
\end{equation}
\end{widetext}

\noindent Here, \( B_{l,m,l',m',l'',m''} \), \( C_{l,m,l',m',l'',m''} \), and \( D_{l,m,l',m',l'',m''} \) are the multipolar mixing coefficients provided in Appendix D of \cite{aly2024nonlinearities}. In this context, we have ignored the source term for the linear gravitational perturbation to simplify the expression. Including these terms would contribute a source component independent of \( \Phi \) or \( \psi \), as well as the linear electromagnetic modes \( \Phi^{(1)} \), which depend on the specifics of the electromagnetic potential.
\vspace{1mm}

\noindent Generically, beyond linear electromagnetic perturbations, at higher orders \( \Phi^{(j)}, j \geq 2 \), gravitational perturbations will similarly source higher-order electromagnetic modes. A similar analysis, as outlined in Section II of \cite{aly2024nonlinearities}, can be applied here. For instance, at second order in electromagnetic QNMs, we expect to observe phenomena that only arise with higher-level mixing between the two fields, as seen in the gravitational case. For example at second order in $\alpha$, if a gravitational linear QNM has a frequency \( \omega^{(1)}_i \) and a linear electromagnetic QNM has a frequency \( \Omega^{(1)}_j \), part of the quadratic electromagnetic spectrum will include frequencies quadratic in both gravitational-electromagnetic modes \( \Omega^{(2)}_i = \Omega^{(1)}_j \pm \omega^{(1)}_k \) as shown in Fig. \ref{fig:myPlot}, however quadratic modes in electromagnetic will be absent.
\vspace{1mm}

\noindent In most astrophysical scenarios, such as gravitational collapse or mergers, gravitational effects are expected to dominate. In this context, searching for the mixing between electromagnetic and gravitational fields suggests that electromagnetic QNMs with gravitational imprints may be more interesting than the reverse, but the richness of typical BNS and other magnetized phenomena will make it extremely hard to resolve the aforementioned effect. Nevertheless, both approaches could serve as tests for minimal coupling within the framework of General Relativity or for more complex couplings in modified gravity theories. For example, if we consider a different coupling such as the Einstein-Maxwell-Dilaton theory \cite{Pacilio_2018, Hirschmann_2018,Ferrari_2001}, the induced QNMs would exhibit different frequencies. Moreover, if we can imagine an astrophysical phenomenon where both effects are pronounced, it could present a valuable opportunity for multi-messenger astronomy\cite{multi_messenger}.


\noindent Following \cite{aly2024nonlinearities}, we briefly consider a sample source term within the Dirac delta potential approximation, detailed in section \ref{Dirac Delta potaioal} and section III of \cite{aly2024nonlinearities}, and solve for the resulting second-order electromagnetic perturbation. If we assume a quadratic QNM source term, as in \cite{Lagos_2023_GreenFunctionAnalysis_Quadratic_Diracdelta}, we have
\begin{equation}
S = \frac{\Phi^{(1)} \Psi^{(1)}}{(1 + \zeta |x|)^2} \propto  \frac{\, e^{-\left( \frac{V_{\text{EM}}}{2} + \frac{V_G}{2} \right) (t - |x|)}}{(1 + \zeta |x|)^2} .
\end{equation}
\noindent Here we only focus on the QNM part. Typically, we should expect a correction to the amplitude of the electromagnetic linear modes and the excitation of new frequencies given by
\[
\Omega^{(2)} = -i \frac{V_G + V_{EM}}{2}.
\]
For \( x > 0 \) and \( x < 0 \), respectively, we have the second-order perturbation \( \Phi^{(2)}_{QNM} \):

\begin{widetext}
\begin{equation}\label{GEM_EM<}
\begin{aligned}
   &\Phi^{(2)}_{QNM}(t,x>0)=\frac{8}{\zeta^2 V_G}e^{- \frac{u V_{\text{EM}}}{2}} 
    \left\{ 
        V_{\text{EM}}e^{- \frac{V_{\text{EM}}}{\zeta} } 
        \left[ 
            \operatorname{Ei} \left( \frac{u V_{\text{EM}}}{2} + \frac{V_{\text{EM}}}{\zeta} \right) 
            - 
            \operatorname{Ei} \left( \frac{V_{\text{EM}}}{\zeta} \right) 
        \right] 
        + \zeta 
    \right\} \\
    & \quad\quad\quad\quad+ 
    \frac{8}{\zeta^2 V_G}  e^{-u\frac{ V_{\text{EM}} + V_G}{2} } 
    \left\{ 
        (V_{\text{EM}} + V_G)  e^{- \frac{V_{\text{EM}} + V_G}{\zeta}} 
        \left[ 
            \operatorname{Ei} \left( \frac{V_{\text{EM}} + V_G}{\zeta} \right) 
            - 
            \operatorname{Ei} \left( (V_{\text{EM}} + V_G)\left(\frac{u}{2}+\frac{1}{\zeta}\right) \right) 
        \right] 
        - \zeta 
    \right\}
\end{aligned}
\end{equation}

\begin{equation}\label{GEM_EM>}
\begin{aligned}
  &\Phi^{(2)}_{QNM}(t,x<0)=  \frac{8}{\zeta^2 V_G} e^{-\frac{v V_{\text{EM}}}{2}} 
    \left\{ 
        e^{-\frac{V_{\text{EM}}}{\zeta}} 
        V_{\text{EM}}\left[ 
             \operatorname{Ei} \left( \frac{u V_{\text{EM}}}{2} + \frac{V_{\text{EM}}}{\zeta} \right) 
            - 
            \operatorname{Ei} \left( \frac{V_{\text{EM}}}{\zeta} \right) 
        \right]
        + \zeta 
    \right\} \\
    & \quad\quad\quad\quad+ 
    \frac{8}{\zeta^2 V_G} e^{-\frac{u V_G}{2} - \frac{v V_{\text{EM}}}{2}} 
    \left\{ 
        (V_{\text{EM}} + V_G) e^{-\frac{V_{\text{EM}} + V_G}{\zeta}} 
        \left[ 
            \operatorname{Ei} \left( \frac{V_{\text{EM}} + V_G}{\zeta} \right) 
            - 
            \operatorname{Ei}\left( (V_{\text{EM}} + V_G)\left(\frac{u}{2}+\frac{1}{\zeta}\right) \right)
        \right] 
        - \zeta 
    \right\}
\end{aligned}
\end{equation}
\end{widetext}

The reader can observe from \eqref{GEM_EM<} and \eqref{GEM_EM>} that the QNM component of the GEM mode in the EM sector has amplitudes that depend on the amplitude of the gravitational potential, while its frequencies are quadratic in both the gravitational and electromagnetic perturbations. Consequently, we anticipate a similar behavior in the case of the full potential: the amplitudes of higher-order EM perturbations will also depend on the details of the gravitational potential, and the QNM frequencies should remain quadratic in both fields.

\section{Perturbations due to point charges}\label{Perturbation Due to Point Charges}
In this section, we analyze the electromagnetic perturbation due to one or more point charges in Schwarzschild coordinates. Consider a particle moving along a worldline parameterized by an affine parameter \(\tau\). The four-velocity vector \(U^\mu\) of the particle is given by
\begin{equation}
U^{\mu} = \left(\frac{dt_p(\tau)}{d\tau}, \frac{d\bar{r}_p(\tau)}{d\tau}, \frac{d\bar{\theta}_p(\tau)}{d\tau}, \frac{d\bar{\phi}_p(\tau)}{d\tau}\right),
\end{equation}
which can also be reparameterized in terms of the time coordinate \(t_p\) as:
\begin{equation*}
U^{\mu} = \frac{1}{\frac{dt_p}{d\tau}} \left(1, \frac{dr_p(t_p)}{dt_p}, \frac{d\theta_p(t_p)}{dt_p}, \frac{d\phi_p(t_p)}{dt_p}\right).
\end{equation*}
The four-current vector \(J_\mu\) associated with the charge \(q\) can then be written as:
\begin{equation}
\begin{aligned}
J_\mu &=\left\{ f(r), -\frac{1}{f(r)}\frac{dr_p(t)}{dt}, -r^2 \frac{d\theta_p(t)}{dt}, -r^2\sin^2\theta \frac{d\phi_p(t)}{dt} \right\} \\
&\quad \times  q \, \frac{ \delta(r - r_p(t))}{r^2} \sum_{lm} Y_{lm}(\theta, \phi) Y^*_{lm}(\theta_p(t), \phi_p(t)).
\end{aligned}
\end{equation}
\noindent This four-current vector can be further decomposed using vector spherical harmonics. The components of the current can be expressed as
\begin{equation}
\begin{aligned}
J_\mu &= \left\{ j_{(1)lm} Y_{lm}, j_{(2)lm} Y_{lm}, j_{(3)lm} \dot{Y}_{lm} + j_{(4)lm} \csc\theta \, Y'_{lm}, \right. \\
&\quad \left. j_{(3)lm} Y'_{lm} - j_{(4)lm} \sin\theta \dot{Y}_{lm} \right\},
\end{aligned}
\end{equation}
where the coefficients \(j_{(i)lm}\) are extracted by projecting over the sphere as outlined in \ref{Vector Harmonics}:
\begin{equation}
j_{(i)lm}=\int d\Omega \mathbf{Y}^{\mu*}_{(i)} J_{\mu} .
\end{equation}
Consequently,
\begin{align}
j_{(1)lm} &= q f(r)\frac{\delta(r - r_p)}{r^2} \, Y^*_{lm}(\theta_p, \phi_p), \\
j_{(2)lm} &= -\frac{q}{f(r)}\frac{dr_p}{dt}\frac{ \delta(r - r_p)}{r^2} Y^*_{lm}(\theta_p, \phi_p), \\
j_{(3)lm} &= \frac{q}{\lambda}\delta\left(r - r_p\right)\left\{im \frac{d \phi_p}{d t} - \frac{d \theta_p}{d t} \mathcal{A}_{lm}[Y_{lm}^*]\right\} , \\
j_{(4)lm} &= \frac{q}{\lambda}\delta\left(r - r_p\right)\left\{im \frac{d \theta_p}{d t} \mathcal{C}_{lm}[Y_{lm}^*] + \frac{d \phi_p}{d t} \mathcal{D}_{lm}[Y_{lm}^*]\right\} .
\end{align}
Here, \(\mathcal{A}_{lm}\), \(\mathcal{C}_{lm}\), and \(\mathcal{D}_{lm}\) are operators that effectively act on the multipolar numbers of the spherical harmonics, inducing mulipolar mode-mixing defined as 
\begin{equation}
\begin{aligned}
\mathcal{A}_{lm}[k_{lm}] &= \sum_{lm} \,k_{lm}\int d\Omega Y_{lm} \frac{\partial Y_{l'm'}^*}{\partial \theta} \,, \\
\mathcal{C}_{lm}[k_{lm}] &= \sum_{lm}\,k_{lm} \int d\Omega \csc \theta \, Y_{lm} Y_{l'm'}^* \,, \\
\mathcal{D}_{lm} [k_{lm}]&= \sum_{lm} \,k_{lm} \int d\Omega \sin \theta \, Y_{lm} \frac{\partial Y_{l'm'}}{\partial \theta} \,.
\end{aligned}
\end{equation}
The effective source term for the Regge-Wheeler equation for multipoles \(l > 0\) (even perturbations) \({}_{e}\Phi\) is given by:
\begin{equation}
{}_{J} \mathcal{S}_{\text{e}\,lm} = \frac{f(r)}{\lambda} \left( \frac{\partial}{\partial r} \left[ r^2 j_{(1)lm} \right] - r^2 \frac{\partial j_{(2)lm}}{\partial t} \right).
\end{equation}
For the point charge case, we obtain:
\begin{equation}
\begin{aligned}
{}_{J} \mathcal{S}_{\text{e}\,lm} =& \frac{q}{\lambda} \Bigg\{  \Bigg[ \, \dot{r}_p(t)\left(\dot{\theta}_{p}(t) \frac{\partial Y_{lm}^*}{\partial \theta} - im\,\dot{\phi}_{p}(t) Y_{lm}^*\right) \\
& + \left(f(r)\frac{f(r)}{dr} + \ddot{r}_p(t)\right) Y_{lm}^* \Bigg] \delta\left(r - r_p(t)\right) \\
& + \left[f(r)^2 - \dot{r}_p^2(t)\right] \delta^{\prime}\left(r - r_p(t)\right) Y_{lm}^* \Bigg\}.
\end{aligned}
\end{equation}
The odd perturbation \({}_{o}\Phi\) is described by
\begin{equation}
{}_{J} \mathcal{S}_{\text{o}\,lm} = - f(r) j_{(4)lm}.
\end{equation}
Subsequently, for the point particle case we have
\begin{equation}
\begin{aligned}
{}_{J} \mathcal{S}_{\text{o}\,lm} = & -\frac{q}{\lambda} f(r) \delta\left(r - r_p(t)\right) \left\{im \frac{d \theta_p}{d t} \mathcal{C}_{lm}[Y_{lm}^*] \right.\\
& \left. + \frac{d \phi_p}{d t} \mathcal{D}_{lm}[Y_{lm}^*]\right\} .
\end{aligned}
\end{equation}

Finally, the general solution for \({}_{e,o}\Phi_{lm}\) is given by
\begin{equation}
{}_{e,o}\Phi_{lm}\left(r^{\prime}, t^{\prime}\right)=\int_{-\infty}^{\infty} d t \int_{2 M}^{\infty} d r \, G_{lm}\left(t, t^{\prime}, r, r^{\prime}\right) {}_{J} \mathcal{S}_{e,o \, lm},
\end{equation}
where \(G_{lm}(t, t^{\prime}, r, r^{\prime})\) represents the Green's function for the Regge-Wheeler master equation with spin \(s = 1\), see Section II in \cite{aly2024nonlinearities} for further details. For completeness, we emphasize that for any signal observed at a specific event \((t, x)\), the Green's functions define a limited domain of support in the \((t', x')\) plane \cite{Hui_2019,szpak2004QNM}.  There are 2D sub-regions in the \((t', x')\) space that contribute to the event at \((t,x)\) through convolution of the Green's function with the source term \(\mathcal{S}_{e,o \, lm}\). In the case of a point particle, this contribution is restricted to a 1D curve in \((t', x')\). Consequently, the problem reduces to a path integral along the worldline of the particle, which can be reparameterized over a segment of the path with respect to any variable that is one-to-one with the affine parameter on that segment.
\vspace{1mm}

\noindent For example, if we stick to $t_p$ as affine parameter, then for odd perturbation \({}_{o} \Phi_{lm}\) the solution becomes
\begin{equation}\label{odd_time_paramter}
\begin{aligned}
{}_{o} \Phi_{lm}(r^{\prime}, t^{\prime})= & -\frac{q}{\lambda} \int_{-\infty}^{\infty} d t_p \, f\left(r_p(t_p)\right)\left\{im \frac{d \theta_p}{d t_p} \mathcal{C}_{lm}[Y_{lm}^*] \right.\\
& \left. + \frac{d \phi_p}{d t_p} \mathcal{D}_{lm}[Y_{lm}^*]\right\} \, G_{lm}\left(r^{\prime}, t^{\prime}, r_p(t_p), t_p\right).
\end{aligned}
\end{equation}
For the even perturbation \({}_{e} \Phi_{lm}\)
\begin{equation}\label{even_time_paramter}
\begin{aligned}
{}_{e} &\Phi_{lm}\left(t^{\prime}, r^{\prime}\right) =  \frac{q}{\lambda} \int_{-\infty}^{\infty} d t_p \Bigg\{ \left[ \ddot{r}_p(t_p) - f\left(r_p(t_p)\right) \frac{d f\left(r_p\right)}{d r_p} \right. \\
& \left. - im \, \frac{d r_p}{d t_p}\frac{d \phi_p}{d t_p} + m \cot\theta_p\, \frac{d r_p}{d t_p}\frac{d \theta_p}{d t_p} \right] Y_{lm}^* \\
& +\left[\sqrt{(l-m)(l+m+1)} e^{i \hat{\phi}} \frac{d r_p}{d t_p}\frac{d \theta_p}{d t_p}\right] Y_{l,m+1}^* \Bigg\} \\
& \times G_{lm}\left(r^{\prime}, t^{\prime}, r_p(t_p), t_p\right) + \frac{q}{\lambda} \int^{\infty}_{-\infty} d t_p \left.\frac{G_{lm}}{d r}\right|_{r=r_p(t_p)} \\
& \times \left[\dot{r}_p(t_p)^2 - f^2\left(r_p(t_p)\right)\right] Y_{lm}^*.
\end{aligned}
\end{equation}
However, as we will see in the upcoming subsections, a well-chosen parameter will depend on the worldline being examined, and possibly on the $(t,x)$ coordinates in which we are aiming to examine the perturbations. A thoughtful choice of this parameter can significantly simplify the complexity of the computation.
\subsection{Ideal dipole}

The linearity of Maxwell's equations extends naturally to the electromagnetic Regge-Wheeler equation. If the total current \(J_\mu\) is the sum of the currents from two particles, i.e.,
\begin{equation}
J_\mu = J_\mu^{\text{\scriptsize particle(1)}} + J_\mu^{\text{\scriptsize particle(2)}},
\end{equation}
then the resulting scalar field \(\Phi\) is the superposition of the fields from each particle:
\begin{equation}
\Phi = \Phi^{\text{\scriptsize particle(1)}} + \Phi^{\text{\scriptsize particle(2)}}.
\end{equation}
Here, \(\Phi^{\text{\scriptsize particle(1)}}\) and \(\Phi^{\text{\scriptsize particle(2)}}\) are the solutions to the Regge-Wheeler equation for each current. For a simple dipole structure, consider a particle with charge \(q\) on worldline \(\gamma_1\) parameterized by \(r_{p1}\), and another with charge \(-q\) on worldline \(\gamma_2\) parameterized by \(r_{p2}\), where \(r_{p2}-r_{p1}\approx \eta + \mathcal{O}(\eta^2)\)\footnote{In case of radial free fall of a radial dipole $\mathcal{O}(\eta^2)$=0 and the expansion truncates at first order correction in $\eta$.}. Thus, the total field for the dipole could simply be written as a variation of a the perturbation due to a positive charge as 
\begin{equation}
{}_{e,o} \Phi^{d} = \delta \, {}_{e,o}\Phi^{+}.
\end{equation}

The complexity of solving this problem can be divided into three parts:
\begin{enumerate}
    \item Solving for the worldines of charges $q$ and $-q$ considering the electromagnetic force between the particles and self-interaction.
    \item Solving for the Green's function $G_{lm}$.
    \item Evaluating the one-dimensional integral to obtain \({}_{e,o} \Phi\).
\end{enumerate}
We will now consider some simple cases.

\subsection{Examples}\label{examples}
\subsubsection{Point charge in radial free fall}
Consider a charge $q$ freely falling radially from rest with $E=1$ and with angular coordinate $(\theta_0,\phi_0)$, while $\frac{dr_p}{dt_p}=-\beta(r_p) f(r_p)$ and $j_{(3)lm}=j_{(4)lm}=0$ vanishes. We have
\begin{equation}
\begin{aligned}
{}_{o} \Phi_{lm}\left(t^{\prime}, r^{\prime}\right) = 0 .
\end{aligned}
\end{equation}
After re-paramterization of the integral in terms of $r_p$, as outlined in Appendix \ref{re-param}, the even part is 
\begin{equation}
\begin{aligned}
{}_{e} \Phi_{lm}\left(t^{\prime}, r^{\prime}\right) = \frac{-q Y_{lm}^*}{\lambda} \Bigg[ & \int d r_p \Bigg\{\left(-\frac{3 f(r_p)}{2 \beta(r_p)} \frac{df(r_p)}{dr_p}\right) G_{lm} \\
& + \left.\frac{d G_{lm}}{d r}\right|_{r=r_p} \left(-\frac{f(r_p)^2}{\beta(r_p)}\right) \Bigg\} \Bigg].
\end{aligned}
\end{equation}

\subsubsection{Point charge in a circular orbit}
Consider a charge \(q\) in a circular orbit at a constant radius \(r_p\), with angular coordinates \((\theta_0, \phi_p(t_p))\). The angular position \(\phi_p(t_p)\) is related to  time \(t_p\) by \(t_p = \frac{\phi_p}{\bar{\Omega}}\), where \(\bar{\Omega}\) is the constant angular velocity of the particle, and \(j_{(2)lm} = 0\). The odd-parity perturbation is given by
\begin{equation}
\begin{aligned}
{}_{o} \Phi_{lm}\left(t^{\prime}, r^{\prime}\right) = \frac{q}{\lambda} i m f(r_0) 
&\int_{\phi_{p i}}^{\phi_{p f}} d \phi_p \, \mathcal{D}_{lm}\left[ Y_{lm}^* \right]_{\theta = \theta_0, \phi = \phi_p}
 \\ 
&\times G_{lm}\left(r^{\prime}, t^{\prime}, r_0, \frac{\phi_p}{\Omega}\right) .
\end{aligned}
\end{equation}

The even-parity perturbation is given by
\begin{equation}
\begin{aligned}
&{}_{e} \Phi_{lm}\left(t^{\prime}, r^{\prime}\right) = \frac{-q}{\lambda \Omega} 
\int_{\phi_{p i}}^{\phi_{p f}} d \phi_p \, Y^{*}_{lm}(\theta_0, \phi_p) \\
&\quad \times \smaller{ \left[ f \left. \frac{\partial f(r)}{\partial r} \right|_{r = r_0} 
G_{lm}(t^{\prime}, r^{\prime}, \frac{\phi_p}{\Omega}, r_0) 
+ f^2 \left. \frac{\partial G_{lm}}{\partial r} \right|_{r = r_0} \right]}
\end{aligned}
\end{equation}
\subsubsection{Quasi-circular orbit}
We can imagine a worldline where the particle's motion is confined to a plane at \(\theta = \frac{\pi}{2}\) while undergoing angular motion in the \(\hat{\phi}\) direction, similar to that of a circular orbit. Simultaneously, the particle experiences radial motion with velocity \(v\), where \(v \ll r \hat{\Omega}\), starting from \(r_{\text{inner}}\) and moving to \(r_{\text{outer}}\). We can utilize integrals (\ref{even_time_paramter},\ref{odd_time_paramter}) with \(\frac{d r_p}{d t_p}\) and \(\frac{d \phi_p}{d t_p}\) as constants, while \(\frac{d \theta_p}{d t_p} = 0\). This approach approximates the quasi-circular worldline of the point particle.
\vspace{1mm}

\section{Dirac delta function potential}\label{Dirac Delta potaioal}
Away from the light ring, the potential diminishes, and solutions become asymptotically outgoing (or ingoing) near the flat region (or horizon), which is approximately true for the Regge-Wheeler potential at large distances. This holds analytically for superlocalized potentials like the Dirac delta function $V_{EM}\delta(x)$\cite{Lagos_2023_GreenFunctionAnalysis_Quadratic_Diracdelta}. The Green’s function\footnote{See \cite{Leaver_On_QNMs_Schwarzschild, Lagos_2023_GreenFunctionAnalysis_Quadratic_Diracdelta, Kokkotas_1999} for a detailed discussion of the flat Green’s function component \(G_F\), the QNM component \(G_{QNM}\), and the branch cut component \(G_B\).} is expressed as:
\begin{equation}
\begin{gathered}
G(t-t', r, r') = G_F(t-t', r, r') + G_Q(t-t', r, r'),
\end{gathered}
\end{equation}
The Green’s function is further written as
\begin{small}
\begin{equation}
 G(T, r, r') = -\frac{1}{2} \left[ \Theta(A) - \Theta(B) \right] - \frac{1}{2} e^{-\frac{V_{EM}}{2}B} \Theta(B),
\end{equation}
\end{small}
where \(\Theta\) is the Heaviside step function. The derivative of the Green’s function with respect to \(r'\) is:
\begin{small}
\begin{equation}
\begin{aligned}
f(r') &\frac{\partial \, G(t-t', r, r'')}{\partial r''} \bigg|_{r''=r'} 
= \frac{1}{2} \left[ s \delta(A) - \tilde{s} \delta(B) \right] \\
&\quad - \frac{\tilde{s}}{2} e^{-\frac{V_{EM}}{2}(B)} 
\left[ \frac{V_{EM}}{2} \Theta(B) - \delta(B) \right],
\end{aligned}
\end{equation}
\end{small}\footnote{Taking derivative of with Dirac delta function distribution should be done under the integration sign, that is why we are not canceling the two contributions of $\delta(B)$, in case higher derivatives are needed.}
Here, \(s(x - x')\) and \(\tilde{s}(x - x_{EM})\) are sign functions, taking \(+\) for positive arguments and \(-\) for negative arguments. The terms \(A\) and \(B\) encode the causal structure, defined as:
\begin{equation}
\begin{aligned}
A &= t - t' - |x - x'|, \\
B &= t - t' - |x| - |x'|,
\end{aligned}
\end{equation}

\subsection{Support regions}
The causal structure becomes more evident and sharper with simpler potentials, such as the Dirac delta potential. In this case, the QNMs component of the Green's function is supported by the past light cone of the point $(0, t - x)$, which is defined by the region $B$ in the $(t', x')$ plane.
\vspace{1mm}

\noindent\noindent Focusing on point sources, a particle moving along a trajectory \( \gamma \) contributes to the QNM signal observed at \( (x, t) \) primarily due to the particle’s interaction with the potential along its worldline. Almost all of this contribution arises from the particle's history in proximity of the potential. Specifically, if the particle starts from an asymptotically far region, $r \rightarrow \infty$, its contribution begins at $t' \rightarrow -\infty$ and ends when its trajectory exits the past light cone of the point $(0, t - x)$ in the $(t', x')$ plane.
\vspace{1mm}

\begin{widetext}

\begin{figure}[ht]
    \centering
    
    \begin{minipage}{0.3\textwidth}
        \centering
        \includegraphics[width=\textwidth]{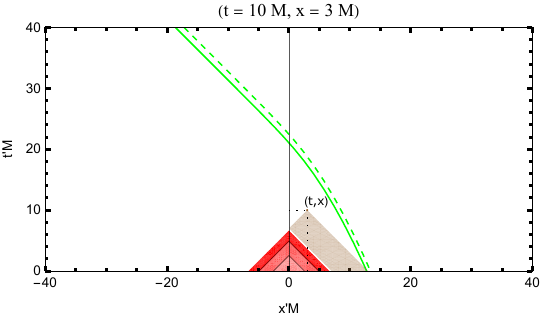}
        \subcaption{}
    \end{minipage}
    \hfill
    \begin{minipage}{0.3\textwidth}
        \centering
        \includegraphics[width=\textwidth]{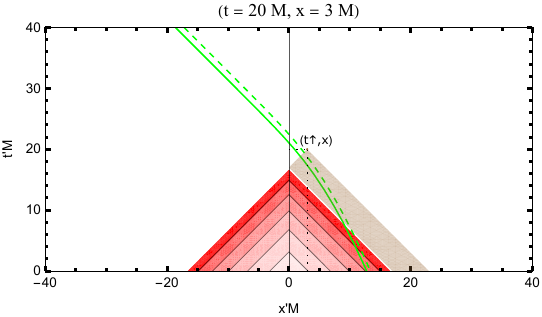}
        \captionsetup{justification=centering}
        \subcaption{}
    \end{minipage}
    \hfill
    \begin{minipage}{0.3\textwidth}
        \centering
        \includegraphics[width=\textwidth]{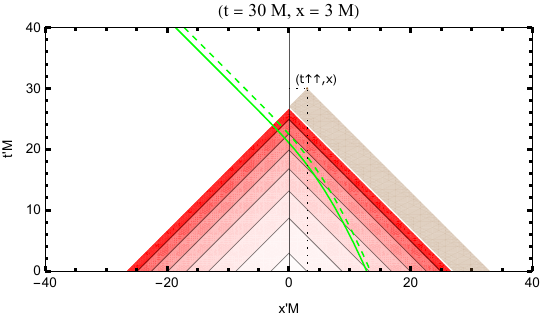}
       \captionsetup{justification=centering}
       \subcaption{}
    \end{minipage}
    
    \vspace{0.4cm} 
    
    \begin{minipage}{0.3\textwidth}
        \centering
        \includegraphics[width=\textwidth]{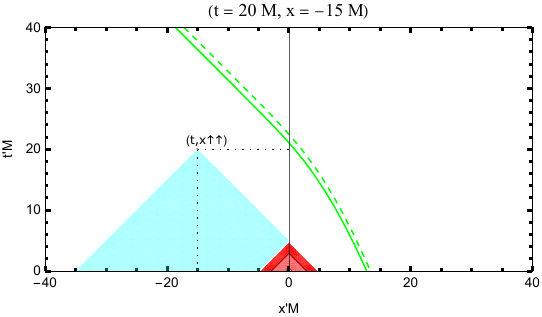}
        \captionsetup{justification=centering}
        \subcaption{}
    \end{minipage}
    \hfill
    \begin{minipage}{0.3\textwidth}
        \centering
        \includegraphics[width=\textwidth]{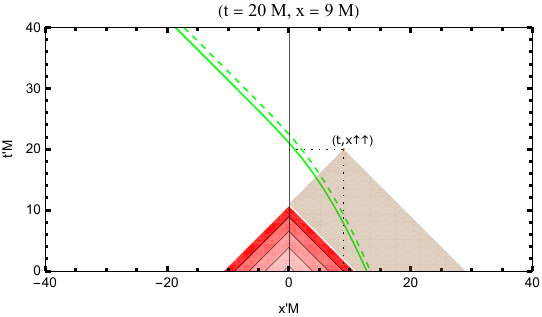}
        \captionsetup{justification=centering}
        \subcaption{}
    \end{minipage}
    \hfill
    \begin{minipage}{0.3\textwidth}
        \centering
        \includegraphics[width=\textwidth]{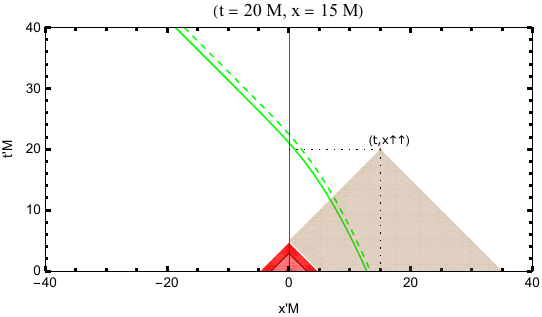}
        \captionsetup{justification=centering}
        \subcaption{}
    \end{minipage}
\captionsetup{format=plain,justification=justified}
 \caption{The red region represents \( G_{Q} \), the QNM support region, with redness indicating the weight of \( G_{Q} \) (where bright red signifies lower weight). For \( x > 0 \), the brown region denotes \( G_{F} \), the flat support region, while for \( x < 0 \), \( G_{F} \) is shown in cyan. Figures (a), (b), and (c) illustrate the expansion of the support regions over time at a fixed observation distance of \( x = 3M \), shown at \( t = 10M \), \( t = 20M \), and \( t = 30M \), respectively. Comparing (b), (e), and (f) highlights the support regions' shrinking effect as \( x \) increases from \( 3M \) to \( 9M \) to \( 15M \) for the same moment in time $20M$. As the observation distance \( x \) increases, the QNM support region becomes more localized around the potential, leading to an asymptotic observer perceiving an almost constant amplitude. Though the amplitude generally depends on both \( t \) and \( x \), this sensitivity diminishes with increased distance as the support region becomes concentrated near the potential. Comparison between (d) and (f) highlights the flat regions \( G_{F} \) at \( x = 15M \) (in brown) and \( x = -15M \) (in cyan). The solid green and dashed green lines represent point charges \( +q \) and \( -q \), separated by \( \eta = 0.6M \), with their initial position at \( r_0 = 12.7M \) and \( r_0+\eta = 13.3M \). The asymmetry in the worldlines of these charges illustrates that, for both flat and QNM parts, for \( x < 0 \) (inside the potential at \( x = 0 \)), contributions are delayed compared to events at the same distance \( x > 0 \) (outside the potential).}
    
    \label{fig:overall_label}
\end{figure}

\end{widetext}

\noindent In many cases, as seen in the literature \cite{Okuzumi_2008,Lagos_2023_GreenFunctionAnalysis_Quadratic_Diracdelta}, the source's contribution is considered to start at a finite time, particularly when gravitational QNMs source the quadratic QNMs, as in binary black hole merger scenarios. While starting the particle's contribution at $t' \rightarrow -\infty$ helps get rid of contribution from electromagnetic field I.C., it can be helpful to consider the particle (the source) starting its contribution at a finite time at some large radius $r_0$, to simplify the analysis. By setting the beginning of time to $t = 0$, the source's contribution is defined by its intersection with the region $B$, bounded below by $t' = 0$ in the $(t', x')$ space. For further clarification, visual references are useful; see Figures \ref{fig:overall_label} (a–f).
\vspace{1mm}

\noindent In the Figures \ref{fig:overall_label}, we imagine a point charge $q$ described by the solid green curve  contributing to the signal observed at $x = 3M$. Initially, in (a), there is no QNM contribution at $t = 10M$, but as time progresses as shown in (b) and (c), the support region $B$ expands, capturing more of the particle's worldline as it enters the region. In contrast, when the observation point $x$ gets further away from the potential (as shown in (f)) the support region shrinks down. In that regime, in the limit of $x\rightarrow\infty$, the contribution only near the potential will excite QNM signal. If we assume that the particle started from $r \rightarrow \infty$ while $t \rightarrow -\infty$, the worldline would always remain inside the region $B$ of any $(t, x)$, and these past light cones $B$ would extend to infinity. 
\vspace{1mm}

\noindent For $G_{Q}$, the redness of the contours, increases on a light-like mesh, with inner past light cones appearing whiter than the outer ones. The outermost part represents the edge of the signal, while the innermost part corresponds to the signal's end. 
\vspace{1mm}

\noindent The integration limits are $(t,x)$-dependent. As we have seen before, if we choose to parameterize the path integral, defined by the overlap of the green curve with the $B$ region (colored in red), using the parameter $r_p$ for certain geodesics\footnote{i.e. radial free fall} the integration limits are $r_{pBmax}(t,x)$ and $r_{pBmin}(t,x)$. For a particle starting at infinity, we generally have $r_{pBmin}(t,x) \rightarrow \infty$. However, solving for $r_{pBmax}(t,x)$ requires solving for $B = 0$ in terms of $r_p$, given by
\[
 t - |x|= t'(r_p) + |x'(r_p)|,
\]
ensuring the integration is properly constrained and accounts for the causal structure encoded in $B$.
\vspace{1mm}

\noindent To study the flat contribution of the Green's function, the situation is analogous except that support region is $A - B$, where $A$ represents another past light cone centered at $(x, t)$. The reader can observe that the flat part's density plot, as in figures \ref{fig:overall_label}, is constant and colored in brown for $x>0$ (or cyan for $x<0$), as the value of $G_{F}$ inside $A - B$ remains a constant $=-1/2$. 
\vspace{1mm}

\noindent Another contribution we should account for is what, mathematically speaking, can be regarded as the evaluation of the derivative of a Dirac delta distribution under the integration sign. Physically, we can interpret this derivative as an additional contribution arising from the edges of the QNM region $B$ and the flat region $A - B$, where the particle is about to penetrate those regions. This contribution should manifest at the edge of the signal. Hence, we will refer only on the integration with $G_{Q}$ as the QNM part.
\vspace{1mm}

\noindent Moreover, the problem is not fully symmetric around the potential peak $x=0$. Although the Green's function itself is symmetric, our particular source term, the worldline of the particle, freezes in proximity to the horizon $x\rightarrow -\infty$. Therefore, only at late times, for equal $\Delta x$ around the potential near the horizon or near the asymptotic flat region, will a QNM be observed. The same holds true for flat contributions, see (d) and (f) for comparison.

\section{First-order electromagnetic perturbation due to point charges}\label{First-order Electromagnetic Perturbation Due to Point Charges}
We are now in a good position to proceed with solving for the electromagnetic perturbation for radial free fall case. Generically, the path integrals in \ref{examples} may present challenges; therefore, the choice between a semi-analytical approximation or a numerical solution depends on which aspects of the solution are of primary interest.

\subsection{Point charge in radial free fall}
For simplicity, we assume homogeneous initial conditions on \({}_{e}\Phi^{(1)}_{lm}(r,t)\):
\begin{equation}
\begin{aligned}
{}_{e}\Phi^{(1)}_{lm}(r,t) &= \frac{-q}{\lambda}  Y_{lm}^{*}(\theta_{0}, \phi_{0})\hat{\Phi}^{(1)}(t,r),\\
\hat{\Phi}^{(1)}(t,r) &= I_{1}(t,r) + I_{2}(t,r),
\end{aligned}
\end{equation}
it’s important to note that \(\hat{\Phi}^{(1)}(t,r)\) in this particular case is independent of the multipole numbers \((l, m)\). The integrals \(I_{1}(t,r)\) and \(I_{2}(t,r)\) are defined as follows\footnote{where $\Theta(t - |x|)$ could be factored out just for clarity, although it should still be embedded in the $B$ integrals regardless}: \begin{align}
I_{1} &= \frac{3}{2} \int_{-\infty}^{\infty} dt' \, G \, f(r_p(t'))^2 \frac{d f(r_p(t'))}{d r_p(t')} \notag \\
&\equiv \frac{3}{4} \bigg[ I_{F1A} - \Theta I_{F1B}  + \Theta I_{Q1B} e^{-\frac{V_{EM}}{2}(t-|x|)} \bigg], \label{eq:I1}
\end{align}
and
\begin{align}
I_{2} &= \int_{-\infty}^{\infty} dt' \, \frac{\partial G}{\partial r'} \bigg|_{r'=r_p(t')} f(r_p(t'))^3 \notag \\
&\equiv -\frac{1}{2} \bigg[ s I_{F2A} -\Theta \tilde{s} \frac{V_{EM}}{2} e^{-\frac{V_{EM}}{2}(t-|x|)}  I_{Q2B} \bigg]. \label{eq:I2}
\end{align}
The ``amplitude" $A(t,x)$ of the QNM part can be expressed as
\[
A(t,x)=\frac{3}{4} I_{Q1B} +\frac{\tilde{s}}{4} V_{EM} I_{Q2B},
\]
However, evaluating $I_{Q1B}$ and $I_{Q2B}$ may not be as straightforward as the other integrals, as the reader can see in Appendix \ref{integrals}. While these integrals can be computed numerically, if we solve for $r_{pBmax}(t,x)$ numerically. For simplicity, we will proceed with I.C. on the trajectory that $t=0$ at $r_{pBmin}=r_{0}$ \footnote{if we are interested in the case starting far from the black hole, i.e. $r_{0} \rightarrow \infty$, this will require that $t$ runs $-\infty<t<\infty$.}, and also that $t$ is positive definite. Since the integration limits are often easier to handle in double null coordinates $u'$ and $v'$, we can proceed with the following transformation of variables in the right half on the $(t', x')$ plane as
\begin{align*}
    v &\mapsto v(r_p) = t(r_p) + x(r_p), \\
    -dv& \, f(1+\beta) = \frac{f}{\beta}dr_p,
\end{align*}
while for the left half we have
\begin{align*}
    u &\mapsto u(r_p) = t(r_p) - x(r_p), \\
    -du &\, f(1-\beta) = \frac{f}{\beta}dr_p.
\end{align*}
The integration limits are a bit tricky as we encounter three different cases:
\begin{enumerate}
    \item If we are interested in the case where the particle starts from a finite position $x'(r_{0})>t-|x|$, then particle never enters the past light cone of $(0,t-|x|)$. In this case, it would have zero QNM contribution for $(t, x)$.
    \item The particle exits the past light cone before reaching the potential peak at $x = 0$. There is no contribution from left half of the light cone.
\begin{equation}  
\begin{gathered}
    r_{pBmin}(t,r)=r_{0} \mapsto v =x(r_0)\\
     r_{pBmax}(t,r) \mapsto v=t-|x|.
\end{gathered}
\end{equation}
\begin{equation}
A_{2}(t,x)= \int_{t-|x|}^{x(r_0)} d v' \, (\beta + 1) f \left(\frac{3}{4} \frac{d f}{d r_p} + \frac{\tilde{1}}{4}  V_{EM} \right) e^{v' \frac{V_{EM}}{2}},
\end{equation}
    \item The particle exits after passing $x=0$ the past light cone. Then right side contribution in $v'$:
\begin{equation}  
\begin{gathered}
    r_{pBmin}(t,r)=r_{0} \mapsto v =x(r_0)\\
    \quad\quad v=t'',
\end{gathered}
\end{equation}
where $t''$ represents the time at which the particle passes the peak of the potential at $x=0$. The left side contribution in $u'$ is
\begin{equation}  
\begin{gathered}
u=t''.\\
r_{pBmax(t,r)} \mapsto v=t-|x|.
\end{gathered}
\end{equation}
Hence, the integral should be divided into two halves before and after $x=0$
\begin{equation}
\begin{aligned}
A_{3}(t,x)&= \int_{t - |x|}^{t''} d u' \, (\beta - 1) f \left(\frac{3}{4} \frac{d f}{d r_p} + \frac{\tilde{s}}{4} V_{EM} \right) e^{u' \frac{V_{EM}}{2}}\\ 
&+ \int_{t''}^{x(r_{0})} d v' \, (\beta + 1) f \left(\frac{3}{4} \frac{d f}{d r_p} + \frac{\tilde{s}}{4}  V_{EM} \right) e^{v' \frac{V_{EM}}{2}} .
\end{aligned}
\end{equation}   
\end{enumerate}
While the integration limits are now straightforward, this has resulted in a complex integrand, as was expected since the worldline is timelike. In other words, because it's not easy to invert the function \( u'(r_p) \) (or \( v'(r_p) \)) analytically. However, for some point $(t,x)$ such that $t \approx |x|$, we can expand around $u=0$\footnote{When $x>0$, we can use $u$, for $x<0$, we use $v$}. Physically we can find the initial behavior of the QNM at some position $x$, however as $t$ increases, higher terms in the expansions need to be considered. A Taylor expansion of the segment of the function of $r_p(u)$ around $u=0$ up to the order $n$ can be integrated as follows
\[
\int P_{n}(u) e^{ u\frac{V_{EM}}{2}}= e^{u \frac{V_{EM}}{2} } \sum_{i=0}^n c_i \sum_{k=0}^i(-1)^k \frac{i!}{(i-k)!} \frac{u^{i-k}}{\left(\frac{V_{EM}}{2}\right)^k} ,
\]
where $c_{i}$ are the coefficient of the expansion
\[
(\beta + 1) f \left(\frac{3}{4} \frac{d f}{d r_p} + \frac{\tilde{s}}{4}  V_{EM} \right)
\]
around $u=0$ that could easily be obtained through applying the chain rule.

\noindent Back to case (2), evaluating the integrand above at the upper limit $x(r_{0})$ will contribute to the QNM part with amplitude that only depends on the I.C. of the worldline. However the lower limit $t-|x|$ will result in a flat contribution which is $(t,x)$ dependent.
\begin{widetext}
\begin{equation}
\begin{aligned}
    &e^{-\frac{V_{EM}}{2}(t - |x|)} A_{2}(t, x) = e^{-\frac{V_{EM}}{2}(t - |x|)} \mathbf{A}_{2}(r_{0})- \sum_{i=0}^n c_i \left( \sum_{k=0}^i (-1)^k \frac{i!}{(i-k)!} \frac{(t - |x|)^{i-k}}{\left( \frac{V_{EM}}{2} \right)^k} \right),\\
    &\mathbf{A}_{2}(r_{0})= e^{x(r_0) \frac{V_{EM}}{2}} 
    \sum_{i=0}^n c_i \left( \sum_{k=0}^i (-1)^k \frac{i!}{(i-k)!} \frac{x(r_0)^{i-k}}{\left( \frac{V_{EM}}{2} \right)^k} \right). \\
\end{aligned}
\end{equation}

\end{widetext}

\noindent The case (3) should be very similar to the case (2), and we will end up with QNM part with amplitude only dependent on the initial conditions for the particle and a flat part polynomial in $t-|x|$. However, the coefficients of expansion will be different.
\begin{widetext}
\begin{equation}
\begin{aligned}
   &e^{-\frac{V_{EM}}{2}(t - |x|)} A_{3}(t, x) = e^{-\frac{V_{EM}}{2}(t - |x|)} \mathbf{A}_{3}(r_{0})- \sum_{i=0}^n f_i \left( \sum_{k=0}^i (-1)^k \frac{i!}{(i-k)!} \frac{(t - |x|)^{i-k}}{\left( \frac{V_{EM}}{2} \right)^k} \right),\\
   &\mathbf{A}_{3}(r_{0})= e^{x(r_0) \frac{V_{EM}}{2}} 
    \sum_{i=0}^n d_i \left( \sum_{k=0}^i (-1)^k \frac{i!}{(i-k)!} \frac{x(r_0)^{i-k}}{\left( \frac{V_{EM}}{2} \right)^k} \right)- e^{t'' \frac{V_{EM}}{2}}  \sum_{i=0}^n d_i \left( \sum_{k=0}^i (-1)^k \frac{i!}{(i-k)!} \frac{t''^{\, i-k}}{\left( \frac{V_{EM}}{2} \right)^k} \right)\\
    &+e^{t'' \frac{V_{EM}}{2}} \sum_{i=0}^n f_i \left( \sum_{k=0}^i (-1)^k \frac{i!}{(i-k)!} \frac{t''^{\, i-k}}{\left( \frac{V_{EM}}{2} \right)^k} \right). \\
\end{aligned}
\end{equation}

\noindent Finally, the Maxwell tensor $F$ describing this perturbation can be written as 
\begin{equation}
 F=F_{\text{monopole}}+{}_{e1} F_{\text{rad}}+{}_{e2}F_{\text{int}} ,
\end{equation}

where the monopole contribution as well as the radiative and ``internal" contribution are expressed in terms of \(\Phi^{(1)}\) and the current. The monopole contribution is simply $F_{\text{monopole}\, tr} = q \frac{\Theta(r - r_{p}(t))}{r^2}$.
The radiative contribution is
\begin{equation}
\begin{aligned}
{}_{e1} F_{\text{rad}}=q
\begin{pmatrix}
0 & ** & ** & ** \\
\left[\delta^2(\theta,\theta_0,\phi,\phi_0) - \frac{1}{4\pi}\right] \frac{\hat{\Phi}^{(1)}}{r^2} & 0 & ** & ** \\
-\frac{\partial G_{S^2}}{\partial \theta} f \frac{\partial \hat{\Phi}^{(1)}}{\partial r} & -\frac{\partial G_{S^2}}{\partial \theta} \frac{1}{f} \frac{\partial \hat{\Phi}^{(1)}}{\partial t} & 0 & 0 \\
-\frac{\partial G_{S^2}}{\partial \phi} f \frac{\partial \hat{\Phi}^{(1)}}{\partial r} & -\frac{\partial G_{S^2}}{\partial \phi} \frac{1}{f} \frac{\partial \hat{\Phi}^{(1)}}{\partial t} & 0 & 0
\end{pmatrix}
\end{aligned} .
\end{equation}
While internal contribution is given by
\begin{equation}
\begin{aligned}
{}_{e2} F_{\text{int}}=q
\begin{pmatrix}
0 & 0 & ** & ** \\
0 & 0 & ** & ** \\
-\frac{\partial G_{S^2}}{\partial \theta} f \delta(r-r_{p}(t)) & -\frac{\partial G_{S^2}}{\partial \theta} \beta \, \delta(r-r_{p}(t)) & 0 & 0 \\
-\frac{\partial G_{S^2}}{\partial \phi} f \delta(r-r_{p}(t)) & -\frac{\partial G_{S^2}}{\partial \phi} \beta \, \delta(r-r_{p}(t))  & 0 & 0
\end{pmatrix}.
\end{aligned}
\end{equation}
\end{widetext}
\subsection{Radial ideal dipole in radial free fall}
For simplicity, we will focus on the scenario of radial free fall, and the radial dipole. Additionally, we will disregard the effects of electromagnetic interactions and self-interactions between the two particles along their worldlines. Instead, we consider two charges, \(q\) and \(-q\), with a  separation \(\eta\) along their radial worldlines into the black hole, starting from rest at large distance positions $r_{0}$ and $r_{0}+\eta$ respectively with \(E=1\). 
\vspace{1mm}

\noindent We can derive expressions for an ideal radial dipole in the limit as $\eta \rightarrow 0$ as a variation of this expression with respect to \(r_{0}\) or equivalently with respect to  \(t''\), denoted as \( \delta A(t,x) \). Focusing on the QNM contribution, we will proceed with similar cases as we had before. In the dipole case (2) only segments of wordline $x_p>0$ contribute to the trajectory

\begin{widetext}

\begin{equation}
\begin{aligned}
   \frac{1}{\eta} \frac{\delta A_{2}}{\delta r_0} = \left[ (\beta + 1) f \left( \frac{3}{4} \frac{d f}{d r_p} + \frac{\tilde{s}}{4} V_{EM} \right) e^{v' \frac{V_{EM}}{2}} \right]_{r = r_0}+ \int_{t - |x|}^{x(r_0)} d v' \, \frac{d}{d r_p} \left[ (\beta + 1) f \left( \frac{3}{4} \frac{d f}{d r_p} + \frac{\tilde{s}}{4} V_{EM} \right) \right] e^{v' \frac{V_{EM}}{2}}.
\end{aligned}
\end{equation}
\begin{equation}
\begin{aligned}
 \frac{1}{\eta}\frac{\delta A_{3}}{\delta r_0}&= \frac{r_{0}^{3/2}}{2(r_{0}-2)}\left[(\beta - 1) f \left(\frac{3}{4} \frac{d f}{d r_p} + \frac{\tilde{s}}{4} V_{EM} \right) e^{u' \frac{V_{EM}}{2}}-(\beta + 1) f \left(\frac{3}{4} \frac{d f}{d r_p} + \frac{\tilde{s}}{4}  V_{EM} \right) e^{v' \frac{V_{EM}}{2}}\right]_{r = r_0}\\
&- f(r_{0})\left[ (\beta + 1) f \left(\frac{3}{4} \frac{d f}{d r_p} + \frac{\tilde{s}}{4}  V_{EM} \right) e^{v' \frac{V_{EM}}{2}}\right]_{r=r_{0}}+\int_{t - |x|}^{t''} d u' \, \frac{d}{d r_p}\left[(\beta - 1) f \left(\frac{3}{4} \frac{d f}{d r_p} + \frac{\tilde{s}}{4} V_{EM}\right) \right] e^{u' \frac{V_{EM}}{2}}\\
&+ \int_{t''}^{x(r_{0})} d v' \, \frac{d}{d r_p} \left[(\beta + 1) f \left(\frac{3}{4} \frac{d f}{d r_p} + \frac{\tilde{s}}{4}  V_{EM} \right)\right] e^{v' \frac{V_{EM}}{2}},
\end{aligned}
\end{equation}   
where $\frac{\delta t''}{\delta r_0}=\frac{r_{0}^{3/2}}{2(r_{0}-2)}\eta$. Hence, the EM perturbation is
\begin{equation}
\begin{aligned}
{}_{e}\Phi^{dip(1)}_{lm}(r,t) &= -\frac{q}{\lambda} Y_{lm}^* \eta \, \hat{\Phi}^{dip(1)}(r,t), \\
\end{aligned}
\end{equation}
where $+$ and $-$ refer to positive and negative charges respectively, and $p=q \eta$, assuming that the scale of change of $r_{pzero}$ is much larger than $p$. In a region outside the dipole, there is no monopole contribution.
The radiative contribution is represented by
\begin{equation}
\begin{aligned}
{}_{e1} F_{\text{rad}} &= -p \begin{pmatrix}
0 & ** & ** & ** \\
\left[\delta^2(\theta,\theta_0,\phi,\phi_0) - \frac{1}{4\pi}\right] \frac{\hat{\Phi}^{dip(1)}(t,r)}{r^2} & 0 & ** & ** \\
-\frac{\partial G_{S^2}}{\partial \theta} f \frac{\partial \hat{\Phi}^{dip(1)}(t,r)}{\partial r} & -\frac{\partial G_{S^2}}{\partial \theta} \frac{1}{f} \frac{\partial \hat{\Phi}^{dip(1)}(t,r)}{\partial t} & 0 & 0 \\
-\frac{\partial G_{S^2}}{\partial \phi} f \frac{\partial \hat{\Phi}^{dip(1)}(t,r)}{\partial r} & -\frac{\partial G_{S^2}}{\partial \phi} \frac{1}{f} \frac{\partial \hat{\Phi}^{dip(1)}(t,r)}{\partial t} & 0 & 0
\end{pmatrix}
\end{aligned} ,
\end{equation}
where $\delta^2(\theta,\theta_0,\phi,\phi_0)$ is the Dirac delta function at the radial line of free fall, while $G_{S^{2}}$ is the angular Green's function on a \( S^{2} \) sphere (the expressions can be found in \cite{Fawzi_EM_idealdipole}). There is also another contribution to the electromagnetic field, localized at the position of the points charges, which is referred as \( {}_{e2}F_{\text{int}} \), given by
\begin{equation}
\begin{aligned}
{}_{e2} F_{\text{int}} &= -p \begin{pmatrix}
0 & 0 & ** & ** \\
0 & 0 & ** & ** \\
-\frac{\partial G_{S^2}}{\partial \theta} f \left[\delta(r-r_{p}(t))-\delta(r-r_{p}(t)-\eta)\right] & -\frac{\partial G_{S^2}}{\partial \theta} \beta \, \left[\delta(r-r_{p}(t))-\delta(r-r_{p}(t)-\eta)\right] & 0 & 0 \\
-\frac{\partial G_{S^2}}{\partial \phi} f \left[\delta(r-r_{p}(t))-\delta(r-r_{p}(t)-\eta)\right] & -\frac{\partial G_{S^2}}{\partial \phi} \beta \, \left[\delta(r-r_{p}(t))-\delta(r-r_{p}(t)-\eta)\right]  & 0 & 0
\end{pmatrix} .
\end{aligned}
\end{equation}
Similar to \( F_{\text{monopole}} \), the term \( {}_{e2} F_{\text{int}} \) vanishes outside the dipole. Consequently, its contribution is ignored when calculating the electromagnetic stress-energy tensor \( T_{EM}^{\mu\nu} \), which will source the gravitational perturbation.
\end{widetext}

\section{Numerical solution}\label{Numerical Solution}

In this section, we discuss the results of numerically integrating the contributions to the QNM and flat electromagnetic perturbations arising from a radially free-falling radial dipole. In the figures, the particle's position is marked by a red dashed vertical line, while the peak of the potential, \( V_{\text{EM}} = \frac{M}{10} \), is also indicated. The dipole initially begins around \( x_0 = 200M \) with a separation \( \eta = 1M \).

\begin{widetext}
    
\begin{figure}[H]
    \centering
    \begin{minipage}{0.3\textwidth}
        \centering
        \includegraphics[width=\textwidth]{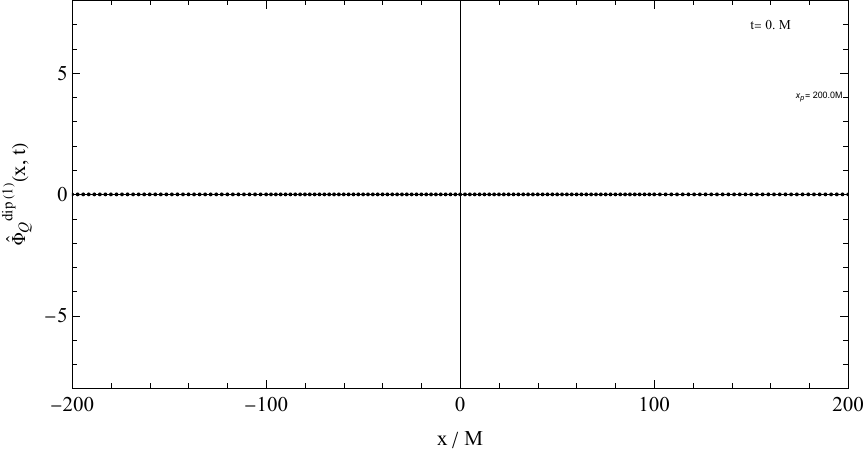}
        \subcaption{}
    \end{minipage}
    \hfill
    \begin{minipage}{0.3\textwidth}
        \centering
        \includegraphics[width=\textwidth]{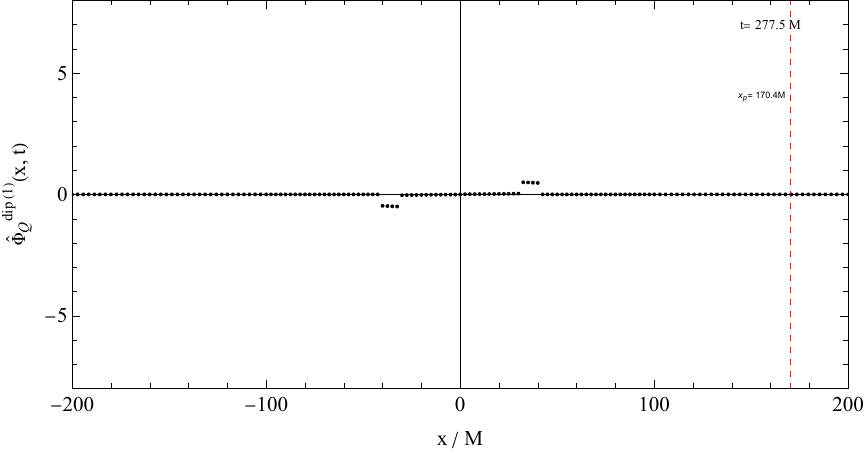}
        \subcaption{}
    \end{minipage}
    \hfill
    \begin{minipage}{0.3\textwidth}
        \centering
        \includegraphics[width=\textwidth]{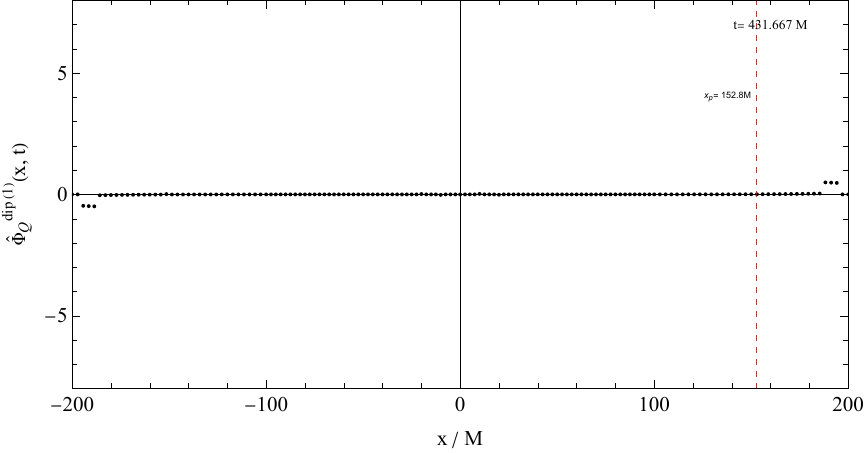}
        \subcaption{}
    \end{minipage}
    
    \vspace{0.4cm} 
    
    \begin{minipage}{0.3\textwidth}
        \centering
        \includegraphics[width=\textwidth]{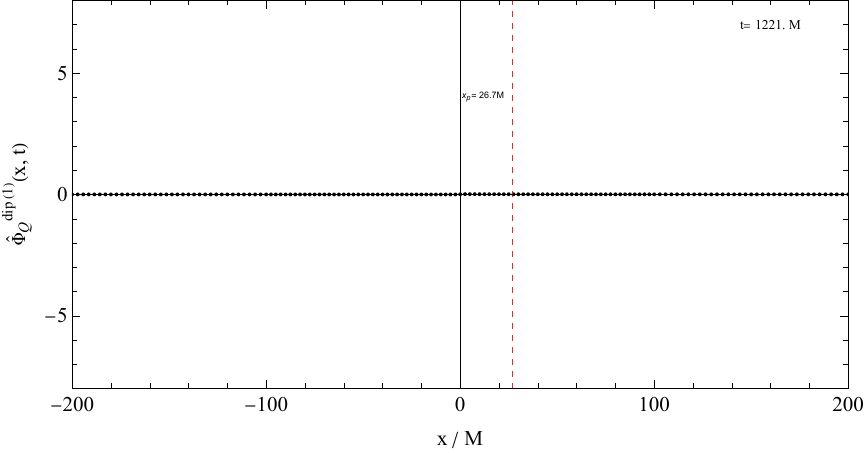}
        \subcaption{}
    \end{minipage}
    \hfill
    \begin{minipage}{0.3\textwidth}
        \centering
        \includegraphics[width=\textwidth]{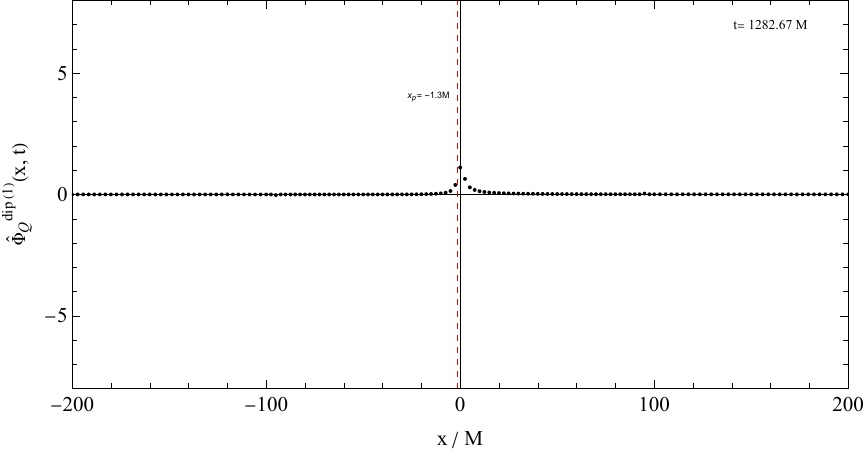}
        \subcaption{}
    \end{minipage}
    \hfill
    \begin{minipage}{0.3\textwidth}
        \centering
        \includegraphics[width=\textwidth]{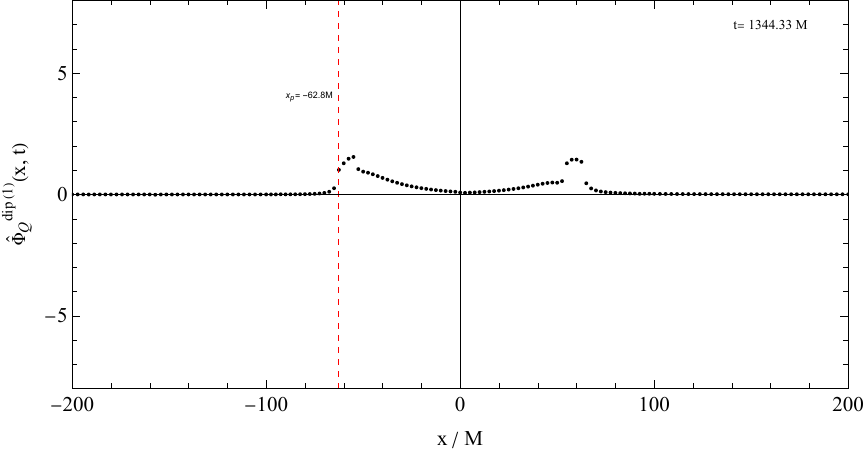}
        \subcaption{}
    \end{minipage}
    
    \vspace{0.4cm} 
    
    \begin{minipage}{0.3\textwidth}
        \centering
        \includegraphics[width=\textwidth]{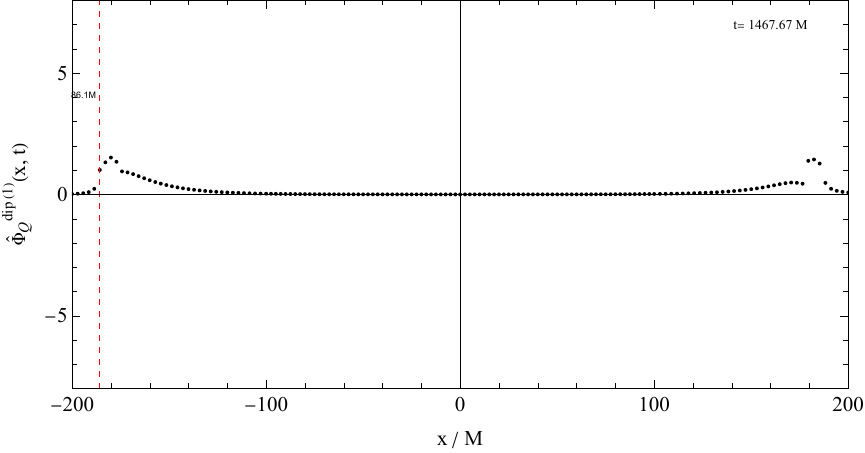}
        \subcaption{}
    \end{minipage}
    \hfill
    \begin{minipage}{0.3\textwidth}
        \centering
        \includegraphics[width=\textwidth]{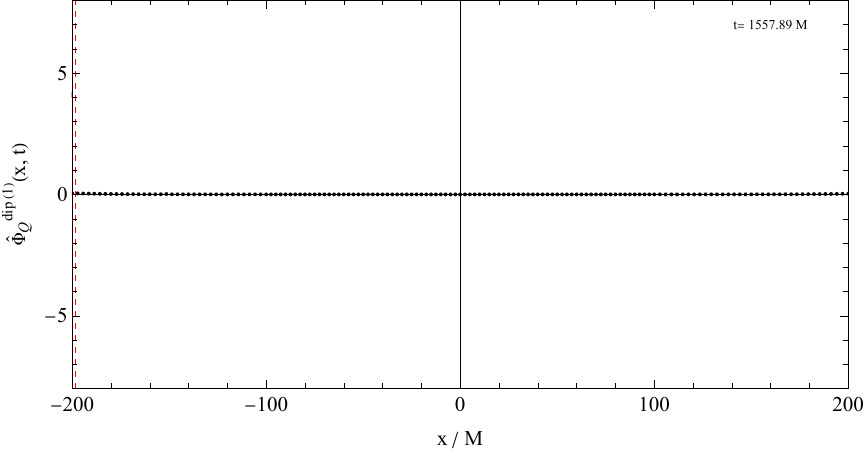}
        \subcaption{}
    \end{minipage}
    \hfill
    \begin{minipage}{0.3\textwidth}
        \centering
        \includegraphics[width=\textwidth]{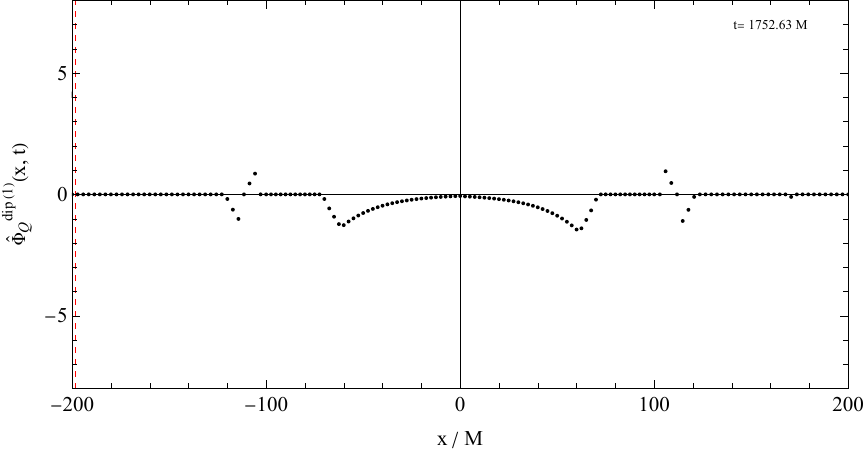}
        \subcaption{}
    \end{minipage}

    \caption{Visual representation of different snapshots of time of the QNM part of the perturbation \( \Phi^{(1)\text{dip}}_{\text{QNM}} \). The \( x \)-range is from \(-200\) to \(200M\), while \( t \) ranges from \( 0 \) to \( 1850M \). }
    \label{fig:9plots}
\end{figure}
\end{widetext}

For the QNM, initially at \( t = 0M \), in Figure \ref{fig:9plots}(a), there is no contribution to \( \hat{\Phi}^{(1)\text{dip}}_{\text{QNM}} \). As time progresses, a constant pulse emerges (see Figure \ref{fig:9plots}(b)), propagating symmetrically from the peak of the potential and moving to the left and right. This event occurs when the particle reaches \( x_p = 175.5M \) at \( t = 231.25M \). This pulse might be related to the fact that charge \( q \) enters the light cone of \( (0, t - x) \), while the other charge, for a short period, remains outside. This highlights two artifacts in this problem: the modeling of the dipole as two discrete charges, and the sharp causality boundaries within the QNM region due to the Dirac delta potential approximation. Additionally, reducing \( \eta \) would control the width of this pulse, while a more realistic, extended potential might reduce its amplitude.

Later, as the particle crosses the potential peak at \( x = 0 \), as shown in (e), around \( t = 1282M \), part of \( \Phi^{(1)\text{dip}}_{\text{QNM}} \) reflects, while another part transmits with nearly equal amplitudes. The transmitted component continues toward the horizon, following the particle's trajectory, as shown in (f) and (g). When analyzed on a logarithmic scale, the slope of \( \log \left[\Phi^{(1)\text{dip}}_{\text{QNM}}\right] \) can be approximately fitted with a line of slope \( -\frac{V_{\text{EM}}}{2} \), suggesting a QNM excitation with an almost constant amplitude, consistent with our expectations.

Around \( t = 1596M \), the particle is nearly frozen at \( r = 2M \), and all QNM contributions to the perturbation vanish completely, as also shown in (h). After some time, around \( t = 1630M \) and \( t = 1670M \), two additional pulses and a shock wave are excited at \( x = 0 \) and begin propagating leftward and rightward, as shown in (i). The second shock wave might suggest that another QNM is excited at late times.

\begin{widetext}

\begin{figure}[H]
    \centering
    \begin{minipage}{0.3\textwidth}
        \centering
        \includegraphics[width=\textwidth]{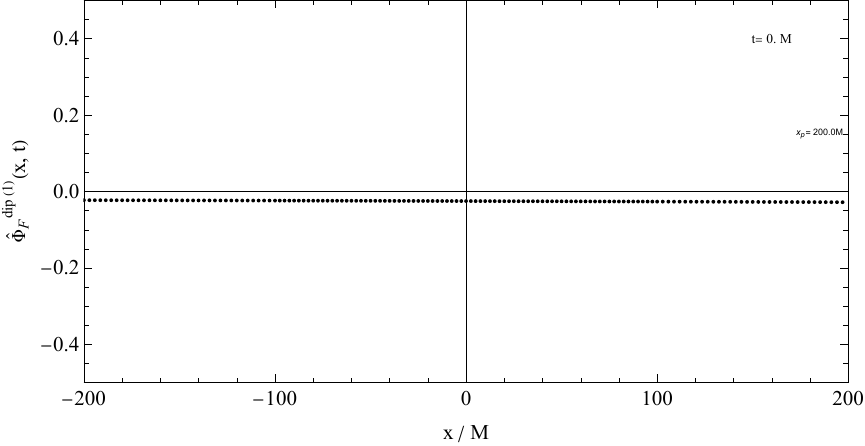}
        \subcaption{}
    \end{minipage}
    \hfill
    \begin{minipage}{0.3\textwidth}
        \centering
        \includegraphics[width=\textwidth]{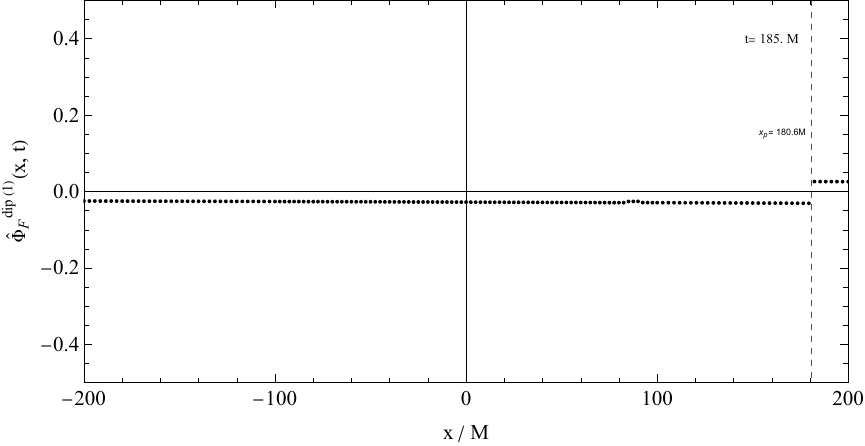}
        \subcaption{}
    \end{minipage}
    \hfill
    \begin{minipage}{0.3\textwidth}
        \centering
        \includegraphics[width=\textwidth]{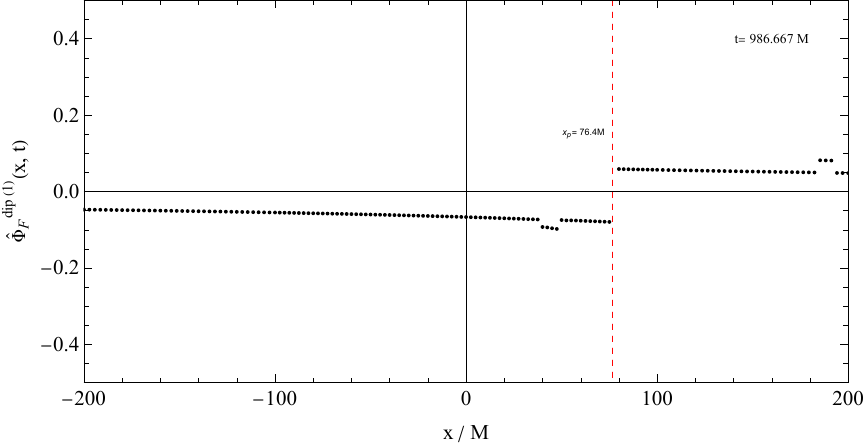}
        \subcaption{}
    \end{minipage}
    
    \vspace{0.4cm} 
    
    \begin{minipage}{0.3\textwidth}
        \centering
        \includegraphics[width=\textwidth]{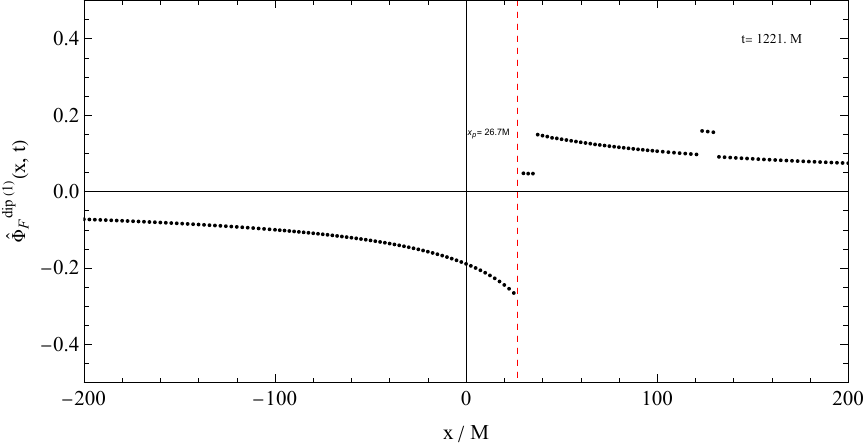}
        \subcaption{}
    \end{minipage}
    \hfill
    \begin{minipage}{0.3\textwidth}
        \centering
        \includegraphics[width=\textwidth]{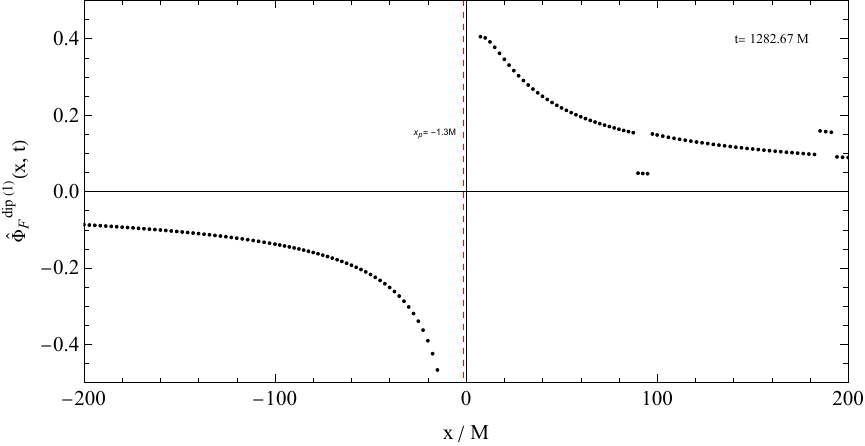}
        \subcaption{}
    \end{minipage}
    \hfill
    \begin{minipage}{0.3\textwidth}
        \centering
        \includegraphics[width=\textwidth]{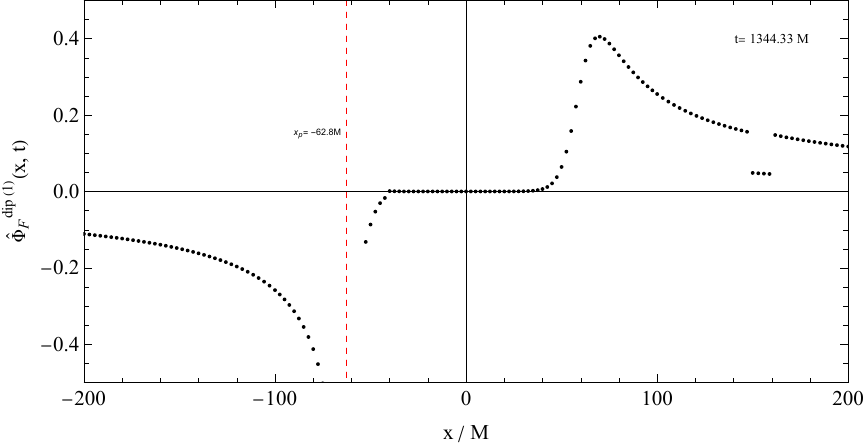}
        \subcaption{}
    \end{minipage}
    
    \vspace{0.4cm} 
    
    \begin{minipage}{0.3\textwidth}
        \centering
        \includegraphics[width=\textwidth]{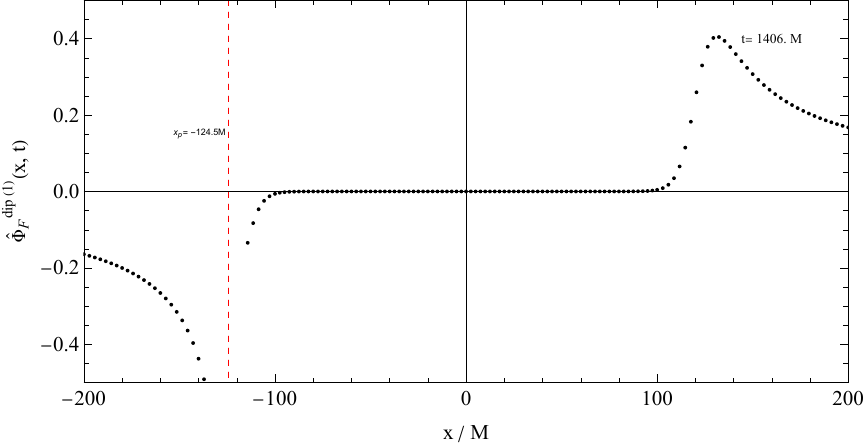}
        \subcaption{}
    \end{minipage}
    \hfill
    \begin{minipage}{0.3\textwidth}
        \centering
        \includegraphics[width=\textwidth]{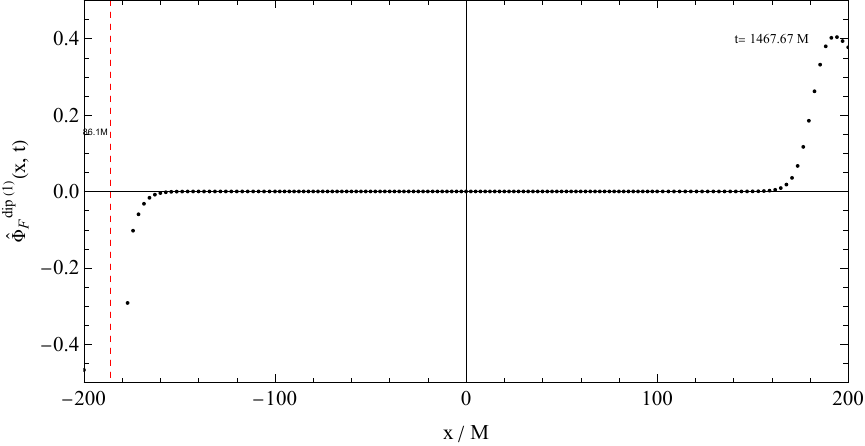}
        \subcaption{}
    \end{minipage}
    \hfill
    \begin{minipage}{0.3\textwidth}
        \centering
        \includegraphics[width=\textwidth]{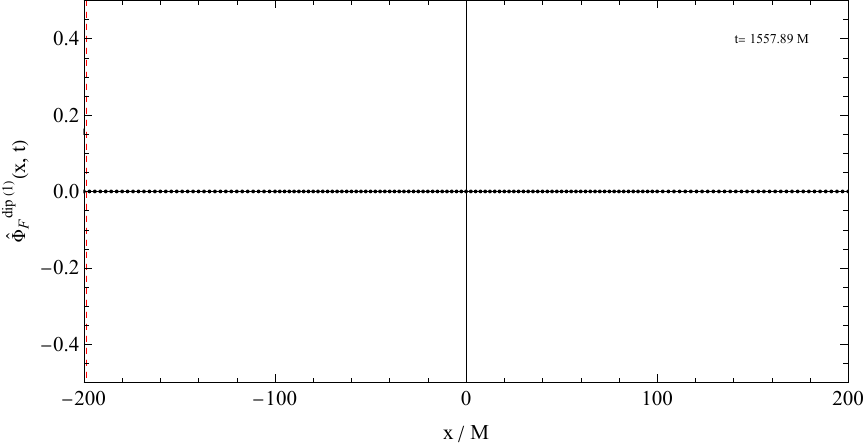}
        \subcaption{}
    \end{minipage}
    \caption{Visual of different snapshots of time of the flat part of the perturbation \( \Phi^{(1)\text{dip}}_{\text{Flat}} \). The \( x \)-range is from \(-200\) to \(200M\), while \( t \) ranges from \( 0 \) to \( 1850M \).  }
    \label{fig:10plots}
\end{figure}

\end{widetext}

For the flat part, as shown in Figure \ref{fig:10plots}(a), (b), and (c), there is a residual field contribution that does not decay and remains constant across the entire spacetime. The field flips sign across the particle, which is expected for a dipole moment. However, this nonlocal behavior is unrealistic, attributable to the fact that we are solving the wave equation in a 1D variable \( x \), while replacing the entire potential with a Dirac delta. This excludes the centrifugal effects, which represent the impact of the other two coordinates. Consequently, if a better approximation were made that preserved the Minkowski limit, we would expect this field to be local and to decay as \( r^{-3} \) away from the potential, as a dipole would in flat spacetime. This suggests that a more accurate approximation could be adopted in future studies if we aim to explore the perturbation more thoroughly.

Furthermore, as the particle approaches the potential, the field builds residual strength near the potential, suggesting that some part of the field will be reflected due to the barrier, as shown in (d) and (e). In (f), we observe a portion of this field being reflected. We anticipate that in a more realistic scenario, as the source of the electromagnetic field moves closer to the potential peak, some of its local (non-radiative) fields would begin to decouple and radiate away, as shown in (g) in our simplified model. Once this shockwave has traveled far enough, all flat contributions vanish, as shown in (i). 

\section{Discussion \& Conclusion  }\label{Conclusion}

 In this work, we have studied the electromagnetic perturbations at first $\Phi^{(1)}$ and second order $\Phi^{(2)}$, using the Regge-Wheeler formalism in the Schwarzschild spacetime. We began by briefly summarizing relevant literature on the first-order electromagnetic perturbations in Schwarzschild coordinates, then reformulated the Faraday tensor $F_{\mu \nu}$ and electromagnetic stress-energy tensor $T_{\mu \nu}$ in terms of the even and odd perturbation scalars ${}_{o,e}\Phi^{(1)}_{lm}$, as well as the current multipoles \( j_{(i) \, lm}\).
\vspace{1mm}

\noindent For the second-order perturbations, we derived the effective source terms for the second-order Regge-Wheeler master equation (with \( s = 1 \)). These sources are quadratic in the first-order gravitational perturbations (both even and odd), given by the Zerilli and Regge-Wheeler master equations (with \( s = 2 \)), and the electromagnetic perturbations (both odd and even). We accounted only for the electromagnetic current \( J \) in these sources, which contributes to induced modes that are linear in gravitational perturbations with an amplitude dependent on the details of the electromagnetic potential. However, for brevity, we did not consider the perturber's effective stress-energy source term \( T^{\mu\nu}_{G} \) in the gravitational perturbations, which could be of some relevance for EMRI studies. We note that including these terms would induce a part of the source in the second-order electromagnetic perturbation that does not depend on first-order perturbations but rather on the physical parameters of the perturber, such as its mass \( M \) and charge \( q \), along with a linear electromagnetic perturbation whose amplitude depends on both electromagnetic and gravitational details.

\noindent Furthermore, by replacing the Regge-Wheeler potential with a Dirac delta function as well as considering a sample source similar to those used in \cite{aly2024nonlinearities, Lagos_2023_GreenFunctionAnalysis_Quadratic_Diracdelta}, and representing a product of linear gravitational and linear electromagnetic perturbations of frequencies \( \omega^{(1)} = -i \frac{V_G}{2} \) and \( \Omega^{(1)} = -i \frac{V_{EM}}{2} \) respectively, we illustrated the mixing that takes place at the second-order electromagnetic QNM \( \Phi^{(2)}_{QNM} \) with a frequency of \( \Omega^{(2)} = -i \frac{V_{EM} + V_G}{2} \).
\vspace{1mm}

\noindent Studying the mixing of the second-order electromagnetic perturbations complements our study of the mixing of second-order gravitational perturbations in \cite{aly2024nonlinearities}. We showed that, when examining the minimally coupled Einstein-Maxwell field equations perturbatively, second-order corrections are induced that are quadratic in the electromagnetic perturbations gravity sector and quadratic in the electromagnetic and gravitational perturbations for electromagnetic sector. Besides its theoretical significance, this mixing could be relevant for multi-messenger astrophysics, particularly for stellar mass and EMRI mergers observed by Advanced LVK and the upcoming LISA. For more discussion on that we refer the reader to \cite{aly2024nonlinearities}.
\vspace{1mm}

\noindent Moreover, we noticed that changing the coupling details, for example by considering the Maxwell-Dilaton-Einstein equations, induces a different mixing spectrum. In this context, this mixing phenomenon is relevant not only for testing the minimal coupling principle but also for probing modified theories of gravity.
\vspace{1mm}

\noindent For the first-order perturbation, we reduced the calculations of the perturbation from a point charge source to a one-dimensional path integral. A suitable choice of parameterization can simplify some of these integrals. By approximating the Regge-Wheeler potential as a Dirac delta function and focusing on two cases --- a single charge \( q \) and a radial dipole moment \( p = q \eta \) --- we found that the flat part of the perturbation is easy to derive analytically, though the QNM part requires a semi-analytical approximation or numerical implementation. We carried out both methods in this work. For the dipole case, based on the numerical solution, we concluded that the QNM perturbation is excited with an almost constant amplitude.
\vspace{1mm}

\noindent Nevertheless, the Dirac delta function approximation presents some limitations, as discussed in the numerical integration section. By approximating the full potential with a Dirac delta, we neglect the centrifugal correction, which encapsulates the Minkowski limit of the field's behavior and effectively reduces the problem to one dimension. In this 1D context, the field does not decay as a function of \( r \), which is unrealistic for a 3D problem. This same issue leads to singular multipole coefficients in the stress-energy tensor reported in Appendix A of \cite{aly2024nonlinearities}. On the other hand, a less trivial approximation would be harder to handle analytically and would not yield simple modes like those with the Dirac delta function. Thus, the choice of approximation should depend on the specific aspects intended for investigation. In our case, as we are interested in mixing, a more realistic potential would be beneficial, and in future work we intend to revisit the mixing problem in Teukolsky equations in the Kerr background, using numerical approach.

\vspace{2mm}

\begin{acknowledgments}
We acknowledge the partial support of D.S. through the U.S. National Science Foundation under Grant No. PHY-2310363. Additionally, we express our gratitude to Macarena Lagos, Ariadna Ribes Metidieri, Ahmed Elshahawy, A.K. Gorbatsievich, and Stanislav Komarov for their insightful discussions. Special thanks to Jacob Fields and Ish Gupta for their invaluable guidance and feedback on this manuscript. We also sincerely appreciate the valuable comments provided by David Radice and Elias Most.
\end{acknowledgments}

\appendix

\section{Green's function on $S^2$}\label{Angular Green Function}
The Green's function $G_{S^2}$ on $S^2$ with a unit radius is 
\begin{equation}
\begin{aligned}
\nabla_{S^2} G_{S^2} &= - Y_{00}(\theta,\phi)Y_{00}^{*}(\theta_0,\phi_0) + \frac{\delta(\theta-\theta_0)\delta(\phi-\phi_0)}{\sin\theta} \\
&= \sum_{l > 0, m} Y_{lm}(\theta,\phi)Y_{lm}^{*}(\theta_0,\phi_0), \\
G_{S^2}&= \sum_{l > 0, m} \frac{-1}{\lambda} Y_{lm}(\theta,\phi)Y_{lm}^{*}(\theta_0,\phi_0)\\
 &= \frac{1}{4\pi}\ln\left[1-\cos\gamma \right]\\
\end{aligned}
\end{equation}
where $ \cos\gamma=\cos\theta \cos\theta_0 + \sin\theta \sin\theta_0 \cos(\phi - \phi_0),$ and $\gamma$ is the angle between the two points $(\theta,\phi)$ and $(\theta_0,\phi_0)$.

\[
\frac{\partial G_{S^2}}{\partial \phi}=-\frac{\sin\theta \sin\theta_0 \sin(\phi - \phi_0)}{4\pi (1 - \cos\gamma)} 
\]

\[
\frac{G_{S^2}}{\partial \theta}=\frac{-\cos\theta_0 \sin\theta + \cos\theta \cos(\phi - \phi_0) \sin\theta_0}{4\pi (1 - \cos\gamma)} 
\]

\section{Integrals}\label{integrals}
Here, \( \epsilon_{r_{pBzero}(t, r)} \) is a binary indicator that equals 1 if \( B \) has a zero at a given \((t,r)\), and 0 otherwise. The quantities \( r_{pA_{\min}}(t, r) \) and \( r_{pB_{\min}}(t, r) \) denote the minimal positive values of \( A \) and \( B \), respectively, while \( r_{pA_{\max}}(t, r) \) and \( r_{pB_{\max}}(t, r) \) represent their corresponding maximal positive values.
\vspace{1mm}

For a given \((t,r)\), if \( A = 0 \) (or \( B = 0 \)), it follows that \( r_{pA_{\min}} = 0 \) (or \( r_{pB_{\min}} = 0 \)). Both \( A \) and \( B \) are monotonic functions with respect to the parameter \( r_p \) along the particle's trajectory for a fixed \((t,r)\) point. The partial derivatives of \( A \) and \( B \) with respect to \( r_p \) are positive definite
\begin{equation}
\frac{\partial A}{\partial r_p} = s \beta(r_p)^2 +\frac{1}{\beta(r_p)}-\frac{s}{\beta(r_p)+1}> 0
\end{equation}

\begin{equation}
\frac{\partial B}{\partial r_p} = -\tilde{s} f(r_p) + \frac{1}{\beta(r_p)} + \frac{\beta(r_p)}{f(r_p)}\left[1 - \tilde{s} \beta(r_p)\right] > 0
\end{equation}

Consequently, the causal structure of the Green's function confines the particle's support to a \((t, r)\) cone along its trajectory. In summary, the problem reduces to finding max, min and zero (if it exists) for $A=0$ and $B=0$ as a function of $(t,r)$, as well as evaluating the  two integrals $I_{Q1B}$ and $I_{Q2B}$ which resemble part of the contribution of the QNM mode.
\begin{widetext}
\begin{equation}
\begin{aligned}
I_{F1A} = \int_{2M}^{\infty} d r_p \frac{f\left(r_p\right)}{\beta\left(r_p\right)} \frac{d f\left(r_p\right)}{d r_p} \Theta(A)
 = \left[ \frac{2}{3}\beta^3 - 2\beta \right] \bigg|_{r_{pA_{\min}}(t,r)}^{r_{pA_{\max}}(t, r)},\\
 \end{aligned}
\end{equation}
\begin{equation}
\begin{aligned}
I_{F1B} = \int_{2M}^{\infty} d r_p \frac{f\left(r_p\right)}{\beta\left(r_p\right)} \frac{d f\left(r_p\right)}{d r_p} \Theta(B)= \left[ \frac{2}{3}\beta^3 - 2\beta \right] \bigg|_{r_{pB_{\min}}(t,r)}^{r_{pB_{\max}}(t, r)},\\
 \end{aligned}
\end{equation}
\begin{equation}
\begin{aligned}
I_{Q1B} = \int_{2M}^{\infty} d r_p \, \frac{f(r_p)}{\beta(r_p)} \frac{d f(r_p)}{d r_p} \, e^{\frac{V_{EM}}{2} \left( t(r_p) + \left| x(r_p) \right| \right)}\Theta(B)= \int_{r_{pB_{\min}}(t, r)}^{r_{pB_{\max}}(t, r)} d r_p \, \frac{f(r_p)}{\beta(r_p)} \frac{d f(r_p)}{d r_p} \, e^{\frac{V_{EM}}{2} \left( t(r_p) + \left| x(r_p) \right| \right)},
\end{aligned}
\end{equation}

\begin{equation}
I_{F 2 A}=\int_{2 M}^{\infty} d r_p \frac{f\left(r_p\right)}{\beta\left(r_p\right)} \delta(A)=\epsilon_{r_{pAzero}(t, r)}\frac{f\left(r_{pAzero}(t, r)\right)}{\beta\left(r_{pAzero}(t, r)\right)} 
\end{equation}

\begin{equation}
I_{Q2B} = \int_{2M}^{\infty} d r_p \frac{f\left(r_p\right)}{\beta\left(r_p\right)} e^{\frac{V_{EM}}{2} \left(t\left(r_p\right) + \left|x\left(r_p\right)\right|\right)} \Theta(B)=\int_{r_{pB_{\min}}(t, r)}^{r_{pB_{\max}}(t, r)} d r_p \frac{f(r_p)}{\beta(r_p)} \, e^{\frac{V_{EM}}{2} \left( t(r_p) + \left| x(r_p) \right| \right)},
\end{equation}

\end{widetext}

\section{Re-parameterization of perturbation integral equation}\label{re-param}
We can parameterize the path integral in terms of $r_p$ as well:
\begin{equation}
\begin{aligned}
{}_{o} \Phi_{lm}(r^{\prime}, t^{\prime})= & -\frac{q}{\lambda} \int_{2 M}^{\infty} d r_p \, f\left(r_p\right)\left\{im \frac{d \hat{\theta}}{d r_p} \mathcal{C}_{lm}[Y_{lm}^*] \right.\\
& \left. + \frac{d \hat{\phi}}{d r_p} \mathcal{D}_{lm}[Y_{lm}^*]\right\} \, G_{lm}\left(r^{\prime}, t^{\prime}, r_p, t\left(r_p\right)\right).
\end{aligned}
\end{equation}

\begin{equation}
\begin{aligned}
{}_{e}  &\Phi_{lm}\left(t^{\prime}, r^{\prime}\right) = \frac{q}{\lambda} \int^{\infty}_{2M} d r_p \Bigg\{ \left[\frac{d \dot{r}_p(r_p)}{d r_p} - \frac{f\left(r_p\right)}{\dot{r}_p\left(r_p\right)} \frac{d f\left(r_p\right)}{d r_p} \right. \\
& \left. - im \, \dot{r}\left(r_p\right) \frac{d \hat{\phi}}{d r_p} + m \cot\hat{\theta}\, \dot{r}\left(r_p\right) \frac{d \hat{\theta}}{d r_p} \right] Y_{lm}^* \\
& +\left[\sqrt{(l-m)(l+m+1)} e^{i \hat{\phi}} \dot{r}\left(r_p\right) \frac{d \hat{\theta}}{d r_p}\right] Y_{l,m+1}^* \Bigg\} \\
& \times G_{lm}\left(r^{\prime}, t^{\prime}, r_p, t\left(r_p\right)\right) \\
& + \frac{q}{\lambda} \int^{\infty}_{2M} d r_p \left.\frac{G_{lm}}{d r}\right|_{r=r_p} \left[\dot{r}_p - \frac{f^2\left(r_p\right)}{\dot{r}_p}\right] Y_{lm}^*.
\end{aligned}
\end{equation}
where \(\hat{\phi}(r_p)=\phi(t(r_p))\) and \(\hat{\theta}(r_p)=\theta(t(r_p))\). This re-parameterization in terms of $r_p$ is valid as long as the particle undergoes radial motion in Schwarzschild coordinates. Still, if the particle passes through the same radial point at different moments in time, the integral needs to be divided into the one-to-on domains and be handled separately. On the other hand, for a purely circular trajectory, we should re-parameterize using the angular coordinate instead.

\section{4D vector harmonics}\label{Vector Harmonics}

In this appendix, spherical decomposition in \cite{KarlMartel_2005_RW} is extended to 4D vector representations, following the same approach of Lousto and Barack \cite{Barack_2005}, but for vectors instead of tensors. The harmonics are defined on the two-dimensional sphere \( S^2 \), where scalar spherical harmonics \( Y_{lm}(\theta,\phi) \) orthogonality are extended to the 4D vector harmonics \( \mathbf{Y}_{\mu \, lm}(\theta,\phi) \). The vector harmonics \( Y^{\mu}_{(1)\, lm}, Y^{\mu}_{(2)\, lm}, Y^{\mu}_{(3)\, lm} \) are categorized as even-parity, while \( Y^{\mu}_{(4)\, lm} \) belongs to the odd-parity class. These are expressed as:
\begin{equation}
\mathbf{Y}^{\mu}_{(1)\, lm} = \left( 
\begin{array}{c}
Y_{lm} \\
0 \\
0 \\
0 
\end{array} 
\right), \quad
\mathbf{Y}^{\mu}_{(2)\, lm} = \left( 
\begin{array}{c}
0 \\
Y_{lm} \\
0 \\
0 
\end{array} 
\right),
\end{equation}

\begin{equation}
\mathbf{Y}^{\mu}_{(3)\, lm} =\frac{1}{\lambda} \left( 
\begin{array}{c}
0 \\
0 \\
\dot{Y}_{lm} \\
\csc^2 \theta Y_{lm}^{\prime} 
\end{array} 
\right), \quad
\mathbf{Y}^{\mu}_{(4)\, lm} = \frac{1}{\lambda} \left( 
\begin{array}{c}
0 \\
0 \\
\csc \theta Y_{lm}^{\prime} \\
-\csc \theta \dot{Y}_{lm}
\end{array} 
\right).
\end{equation}
where \(\dot{Y}_{lm}\) and \(Y_{lm}^{\prime}\) represent the derivatives with respect to \(\theta\) and \(\phi\), respectively. Their is $l$ and $m$  sub-index label  on the vector harmonics \(\mathbf{Y}^{\mu}_{(i)\, lm}\), but suppressed for simplicity. Using the following two identities: 
\begin{equation}
\int_{S^2} d\Omega \csc\theta \,(Y^{*\prime}_{lm} \, \dot{Y}_{lm} - Y^{\prime}_{lm} \, \dot{Y}^{*}_{lm})  = 0,
\end{equation}
and
\begin{equation}
\int_{S^2} d\Omega  (\csc^2 \theta \, Y^{\prime}_{lm} \, Y^{*\prime}_{lm} + \dot{Y}_{lm} \, \dot{Y}^{*}_{lm}) = \delta_{ll'}\delta_{mm'}\lambda.
\end{equation}
The vector harmonics orthogonality condition is given by:
\begin{equation}
\int_{S^2} \mathbf{Y}^{\mu}_{(i)lm} \mathbf{Y}_{\mu(j)l'm'} \, d\Omega = \delta_{ij} \delta_{ll'} \delta_{mm'} N_{i}(r),
\end{equation}
where \( \sqrt{N_{i}(r)} \) represents the norm of the \(i\)-th vector harmonic.
The components of the even-parity vectors \({}_{e}\mathbf{J}\) and \({}_{e}\mathbf{A}\) are expressed as:
\begin{equation}
{}_{e} \mathbf{J}_{\mu} = 
\begin{pmatrix}
j_1 Y \\
j_2 Y \\
j_3 \dot{Y} \\
j_3 Y^{\prime}
\end{pmatrix}
\quad 
{}_{e} \mathbf{A}_{\mu} = 
\begin{pmatrix}
g Y \\
h Y \\
k \dot{Y} \\
k Y^{\prime}
\end{pmatrix}
\end{equation}
Similarly, the components of the odd-parity vectors \({}_{o} \mathbf{A}_{\mu}\) and \({}_{o} \mathbf{J}_{\mu}\) are given by:
\begin{equation}
{}_{o} \mathbf{J}_{\mu} = 
\begin{pmatrix}
0 \\
0 \\
j_{4} \csc \theta Y^{\prime} \\
-j_{4} \sin \theta \, \dot{Y}
\end{pmatrix}
\quad 
{}_{o} \mathbf{A}_{\mu} = 
\begin{pmatrix}
0 \\
0 \\
a \csc \theta Y^{\prime} \\
-a \sin \theta \, \dot{Y}.
\end{pmatrix}
\end{equation}
The summation over the angular \(l\) and azimuthal \(m\) numbers is implied but not explicitly shown. The functions \(j_{i}, a, g, k\) are dependent on the coordinates \( (t, r) \) and the sub-index $lm$.

\bibliography{main}  
\end{document}